\documentclass[preprint]{aastex}
\def\simgr{\,\hbox{\hbox{$ > $}\kern -0.8em \lower 1.0ex\hbox{$\sim$}}\,}
\def\simle{\,\hbox{\hbox{$ < $}\kern -0.8em \lower 1.0ex\hbox{$\sim$}}\,}

\shortauthors{THORSTENSEN et al.}
\shorttitle{Catalina Cataclysmics}
\begin{document}
\title{Spectroscopy and Photometry of Cataclysmic Variable Candidates
from the Catalina Real Time Survey
\footnote{Based on observations obtained at the MDM Observatory, operated by
Dartmouth College, Columbia University, Ohio State University, 
Ohio University, and the University of Michigan.}
}

\author{John R. Thorstensen and Julie N. Skinner}
\affil{Department of Physics and Astronomy\\
6127 Wilder Laboratory, Dartmouth College\\
Hanover, NH 03755-3528}

\begin{abstract}
The Catalina Real Time Survey (CRTS) has found over 500 cataclysmic
variable (CV) candidates, most of which were previously unknown.
We report here on followup
spectroscopy of 36 of the brighter objects.  Nearly all the spectra
are typical of CVs at minimum light.  One object appears to be
a flare star, while another has a spectrum consistent with a CV but lies,
intriguingly, at the center of a small nebulosity.  
We measured orbital periods for eight of the CVs, and estimated
distances for two based on the spectra of their secondary stars.
In addition to the spectra, we 
obtained direct imaging for an overlapping sample of 37 objects,
for which we give magnitudes and colors.
Most of our new orbital periods are shortward of the so-called 
period gap from roughly 2 to 3 hours.  By considering the 
cross-identifications between the Catalina objects and other
catalogs such as the Sloan Digital Sky Survey, we argue that a
large number of cataclysmic variables remain uncatalogued.
By comparing the CRTS sample to lists of previously-known 
CVs that CRTS does not recover, we find that the CRTS is biased 
toward large outburst amplitudes (and hence shorter orbital periods).
We speculate that this is a consequence of the survey cadence.

\end{abstract}


\section{Introduction}

In cataclysmic variable stars (CVs), a white dwarf primary star accretes 
matter by way of Roche lobe overflow from a binary companion, 
which resembles a main-sequence star.   The variety of 
CV behaviors leads to a complicated taxonomy \citep{warner95}.
Many CVs undergo
dwarf nova outbursts, thought to be caused by
accretion disk instabilities which greatly increase the rate at which
matter moves inward in the disk.  Other CVs, called novalike
variables, remain in a bright state for years at a time.
In still others, called AM Her stars or polars, the matter
that is transferred becomes entrained in a strong white-dwarf
magnetic field, and is funneled directly onto the 
white dwarf's magnetic pole.  

The main driver of CV evolution is thought to be a gradual
loss of orbital angular momentum. This causes the Roche critical
lobe of the secondary star to shrink, leading to a shortening
of the orbital period $P_{\rm orb}$, and driving mass
transfer on long (Gyr) timescales.  The
mechanisms by which angular momentum is lost are not fully
understood.  It is often supposed that magnetic  
braking of the secondary star predominates at longer
periods ($>$ 3 hr), and that magnetic braking
becomes inefficient at short period, so that 
gravitational radiation predominates.
Around $P_{\rm orb}$ = 70 min, the
secondary becomes degenerate and its radius begins to {\it increase}
with mass, leading to a slow {\it increase} in the orbital
radius.  This turnaround is often called the {\it period bounce},
even though it is thought to take place very slowly.  

The histogram of CV orbital periods shows a significant dip
at roughly 2 hr $< P_{\rm orb} <$ 3 hr, known as the {\it gap}
\citep{kolbstillthere}.  This is often explained as follows.
As the secondary loses mass, its thermal timescale increases
to become comparable to the time for the orbit to evolve, with the
result that the secondary exceeds its equilibrium radius.
At about three hours, the secondary becomes fully convective,
reducing the efficiency of magnetic braking.  As the orbital
evolution slows, the secondary detaches from its Roche lobe,
shutting down mass transfer. The detached system continues
to evolve to shorter periods, crossing the gap and eventually
re-establishing contact with the Roche lobe near
$P_{\rm orb} = 2$ hr.  While the mechanism by which this
happens is somewhat speculative, \citet{kniggedonor} shows
that a discontinuity in the secondary stars' radii occurs 
across the gap.

In a steady state, the number of stars with a given $P_{\rm orb}$
should be inversely proportional (roughly) to $\dot P$, the
rate at which the period changes.  If
$\dot P$ really is very slow at short periods, then there
should be a large population of short-period CVs, and the
gradual turnaround at the period bounce around 70 minutes should
lead to a `spike' in the distribution 
\citep{pattlate,barker03,unveils}.

Efforts to confront theories such as this with observation
have often been frustrated, because the sample of known 
CVs is incomplete in ways that are difficult to 
quantify. The discovery
channels for CVs include the following:
(1) Dwarf nova outbursts are conspicuous -- they last for days or 
weeks and typically have amplitudes of several magnitudes.  
(2) Nearly all CVs have unusual colors compared to normal stars,
most conspicuously ultraviolet excesses arising from accretion
processes or (in some cases) the underlying white dwarf.
(3) With the exception of some novalike variables and dwarf novae near the 
peak of outburst, nearly all CVs show emission lines, especially
in the Balmer sequence; these can be strong
enough to be noticed in surveys such as IPHAS \citep{withamiphas}.  
(4) A great many other CVs have been discovered as optical counterparts 
of X-ray sources.  (5) CVs at minimum light can be rather faint
($M > 10$), and a small handful have turned up in proper
motion surveys.

New, large samples of CVs with consistent selection
criteria are potentially useful for
clarifying issues such as the space density and orbital
period distribution of these objects.  Because the colors
of CVs overlap those of quasars, the Sloan Digital Sky Survey 
(SDSS) turned up a large number of spectroscopically confirmed
CVs (\citealt{szkodyi,szkodyii,szkodyiii,szkodyiv,szkodyv,
szkodyvi,szkodyvii,szkodyviii}; hereafter referred to as
SzkodyI-VIII).
\citet{unveils} compiled orbital periods for 137 
of these; with the SDSS sample, they were
finally able to discern the long-predicted period spike. 

Like the SDSS, the Catalina Real Time Survey \citep[CRTS; ][]{drakecrtts} 
has discovered many new CVs.  While the
SDSS CVs were originally selected by color -- 
they were chosen for spectroscopy largely because their
colors overlap those of quasars -- the CRTS 
selects entirely by variability.  Briefly, the CRTS
surveys most of the accessible sky at Galactic latitudes 
$|b| > 10^{\circ}$ and declination $\delta > -30^{\circ}$ 
every lunation, using a 0.7 m Schmidt telescope 
in the Catalina mountains near Tucson, Arizona.  
They search for variability using
a master catalog that reaches to $m \sim 22$.
Objects that show abrupt outbursts of $> 2$ mag amplitude
lasting less than few weeks
are classified as likely CVs, making the CRTS a
prolific source of dwarf novae in particular.

The CRTS maintains a catalog of ``Confirmed/Likely''
CVs on the World Wide Web
\footnote{available at http://nesssi.cacr.caltech.edu/catalina/AllCV.html}.
We downloaded this catalog on 2012 March 7, when it 
contained 584 objects, and used this data set for the
present analysis.  

The CRTS CV sample is of great interest because of its 
depth and selection criteria, so it is a natural
choice for follow-up studies similar to 
\citet{unveils}.  \citet{ww12} describe high-speed
photometry of 20 objects, mostly from the 
CRTS, and give orbital periods for 15, including
two eclipsing dwarf novae and two superhumpers. 
Only two of their sample had periods longer than the
2-3 hour period gap.  The period distribution of their
sample also showed the spike just above the turnaround.
In addition, they found that dwarf novae with more
recorded outbursts tended to have longer orbital
periods than those with fewer outbursts. 

Here, we report on followup spectroscopy and photometry
of the CRTS CVs listed in Table \ref{tab:star_info}.
Most of the CRTS CVs are too faint for us to follow up,
so we selected this sample based largely on apparent
brightness.  Operational considerations, such as the 
ease with which observations could be interleaved
other programs, the nearness of the target
to opposition at midnight, and (for radial velocity targets)
the strength and tractability of emission and absorption
spectra also entered into target selection. 

We find spectroscopic periods for 8 objects, one of which
was independently measured by \citet{ww12}.  In 
addition, we obtained exploratory spectra of 
28 others, largely to assess their feasibility for
radial-velocity studies.  We also obtained 
standardized magnitudes for 37 objects (17 of which
also have spectra).  

{\it Plan of this paper.}  
We describe our equipment and techniques in 
Section \ref{sec:techniques}.   
Table \ref{tab:spectro_summary} summarizes all of our
spectroscopy. In 
Section \ref{sec:explore} we describe the spectra of the 28
objects for which we have only a only a quick spectrum.
The eight objects for which we have radial velocity time series are 
discussed in Section \ref{sec:rvs}; 
Table \ref{tab:velocities} gives the velocities, and 
Table \ref{tab:parameters} lists parameters for
sinusoidal fits to the velocity time series.
Table \ref{tab:photometry} gives magnitudes and colors
of objects for which we have standardized photometry.
Finally, in Section \ref{sec:statistics}
we consider how the CRTS CV samples overlaps with other
lists of CVs, discuss the apparent selection biases 
of CRTS and the implications for the CV population 
in general.

\section{Equipment and Techniques}
\label{sec:techniques}

All our data are from MDM Observatory on Kitt Peak,
Ariznoa.  Nearly all are from the 2.4m Hiltner telescope,
but a single photometric measurement was 
kindly taken by J. Halpern at the 1.3m McGraw-Hill telescope.

The spectroscopic observing setups were as follows.

{\it Modular spectrograph.} For most of the spectra
we used the modular spectrograph and a 600 line mm$^{-1}$
grating.  The detector was either `Templeton', a 
1024$^2$ thinned SITe CCD that gave 2 \AA\ pixel$^{-1}$ 
from 4600 to 6700 \AA , or `Nellie', a thick 
2048$^2$ CCD that gave 1.7 \AA\ pixel$^{-1}$ from 
4460 to 7770 \AA , with vignetting toward the ends 
of the range.  The choice of detector was 
dictated by availability during a series of controller upgrades.
With the modspec, targets are centered using the 
image reflected from the polished slit jaws.  For this,
we used new (2010) slit viewing optics and a 
self-contained Andor Ikon CCD camera unit; this has
greatly improved the acquisition of faint targets with
this instrument.  

All of our radial-velocity time series were taken with 
the modular spectograph.  For wavelength calibration we
obtained comparison lamps in twilight and used the 
night-sky $\lambda$ 5577 line to track spectrograph
flexure during the night.  We also observed flux standards
in twilight when the sky was clear, and bright
O and B-type stars in order to correct approximately
for telluric absorption.  

{\it Ohio State Multi-Object Spectrograph (OSMOS).}  
This versatile instrument \citep{martini11} images the focal plane onto
a 4 k $\times$ 4 k CCD, at a scale of 0.273 arcsec 
pixel$^{-1}$.  Filters can be placed in the parallel
beam of the reducing camera for direct imaging, or 
volume-phase-holographic grism disperser 
can be inserted for spectroscopy.  In spectroscopic
mode, one can insert slits in two different locations
in the focal plane, which yield different wavelength
coverage;
we used the `inner' slit, which gives coverage from 
3960 to 6875 \AA\ at 0.75 \AA\ pixel$^{-1}$.  OSMOS 
has no slit-viewing camera, so targets are acquired
by taking a direct image (without slit and disperser)
and then moving the telescope to align the target with 
the known location of the slit.  We inserted
a $V$ filter for the acquisition exposures, and 
therefore obtained a rough $V$ magnitude (without 
a color transformation) for our targets.
Some humid weather
in 2011 September intermittently caused 
condensation in the center of the detector window; 
fortunately, the 
important H$\alpha$ feature was outside the 
affected region.

For most of the spectroscopic reduction we used
IRAF\footnote{
IRAF is distributed by the National Optical Astronomy 
Observatory, which is operated by the Association of 
Universities for Research in Astronomy (AURA) under 
cooperative agreement with the National Science Foundation.}
routines.  To extract one-dimensional spectra from the 
modspec images, we used a local implementation of the
optimal-extraction algorithm of \citet{horne}.

We computed synthetic $V$ magnitudes for our spectra
using the passband tabulated by \citet{bessell}.  Our
slit was usually 1.1 arcsec wide, which means that
an unknown fraction of the light was lost; also,
many of our spectra were taken through thin cloud.  
Experience suggests that our synthetic magnitudes
are accurate to $\sim 0.2$ mag in good conditions; 
in poor conditions, they can be considerably too faint.

We selected eight of our targets for time-series
radial velocity observations, with the aim of determining
orbital periods.  For these, we pushed observations to large
hour angle in order to avoid daily cycle count aliases.
To measure emission-line radial velocities we used the
convolution algorithms described by \citet{sy80}, in 
which an antisymmetric function is convolved with the line
profile and the zero of the convolution is taken as the 
line center.   We used methods described by \citet{shafter83} to
tune the convolution function for maximum signal.  We estimated
uncertainties in the velocities by propagating the 
counting-statistics errors in the spectral channels; these
estimates do not include possible systematic effects.
For absorption-line velocities, we used the \citet{tondav} 
convolution algorithms as implemented by \citet{kurtzmink}.
To search for periods we used a `residual-gram' method
described by \citet{tpst}.  Once we had established
a period, we fit the time series with sinusoids
of the form $v(t) = \gamma + K \sin[2 \pi(t - T_0) / P]$.
Uncertainties in the fit parameters were estimated from the
scatter around the best fit using the procedure described
by \citet{cash79}.

For direct imaging, we mostly used the `Nellie' CCD,
which gave 0.24 \AA\ per pixel.  This detector is 
insensitive in the ultraviolet, so we used
$BVRI$ filters (although $R$ and 
$B$ were sometimes skipped).  We derived photometric 
transformations from observations of \citet{landolt92} 
standard stars.  The scatter in the transformations
was  generally $< 0.05$ mag.  As noted earlier, we also
have some direct images from OSMOS.

\section{Exploratory Spectra}
\label{sec:explore}

\begin{figure}
\plotone{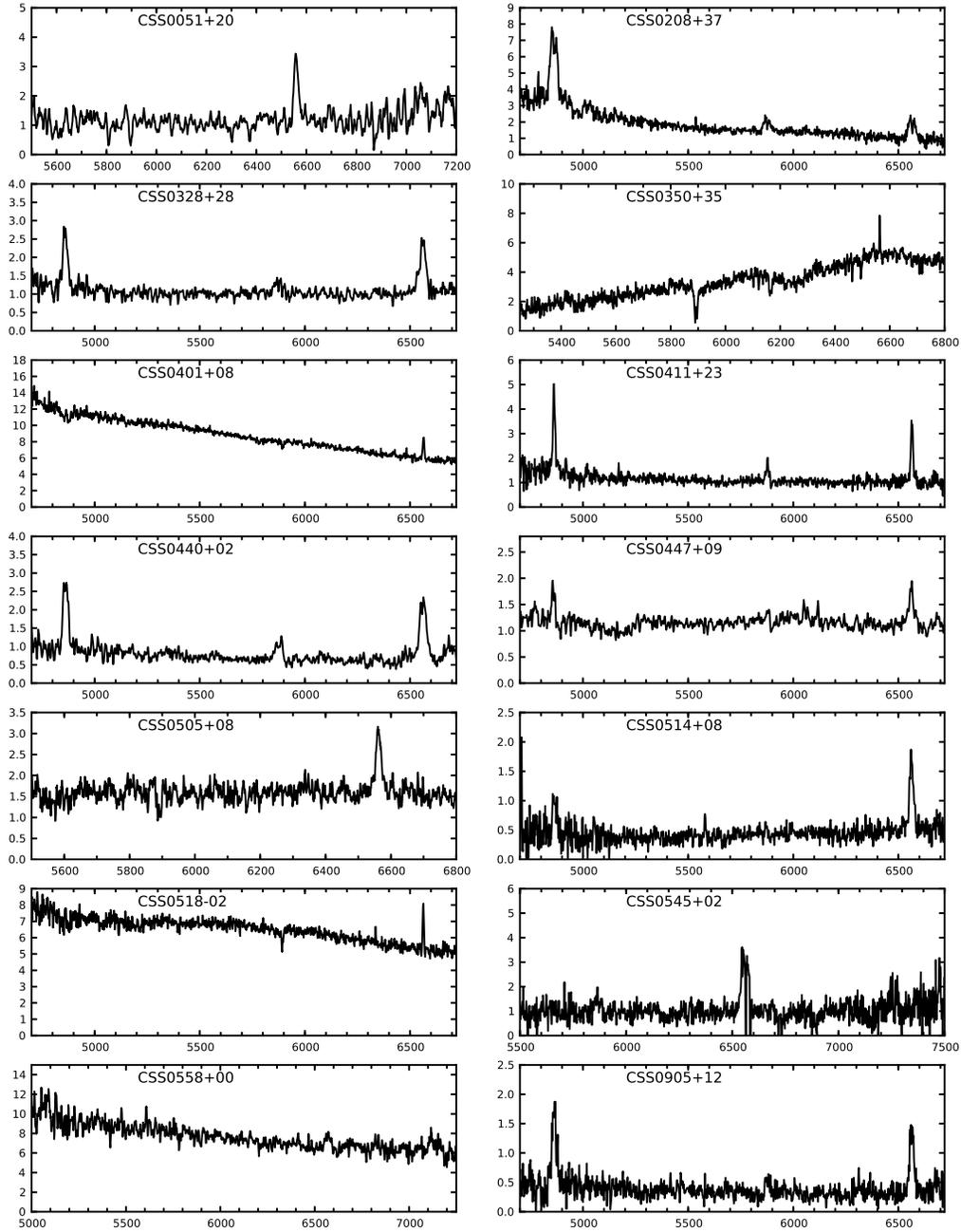}
\caption{
Reconnaiassance
spectra of 14 CRTS CV candidates, arranged by right ascension.  
The vertical axes are $F_\lambda$, in units
of $10^{-16}$ erg cm$^{-2}$ s$^{-1}$ \AA$^{-1}$, while the horizontal
scales give the wavelength in \AA .
}
\label{fig:idspecs1}
\end{figure}

\begin{figure}
\plotone{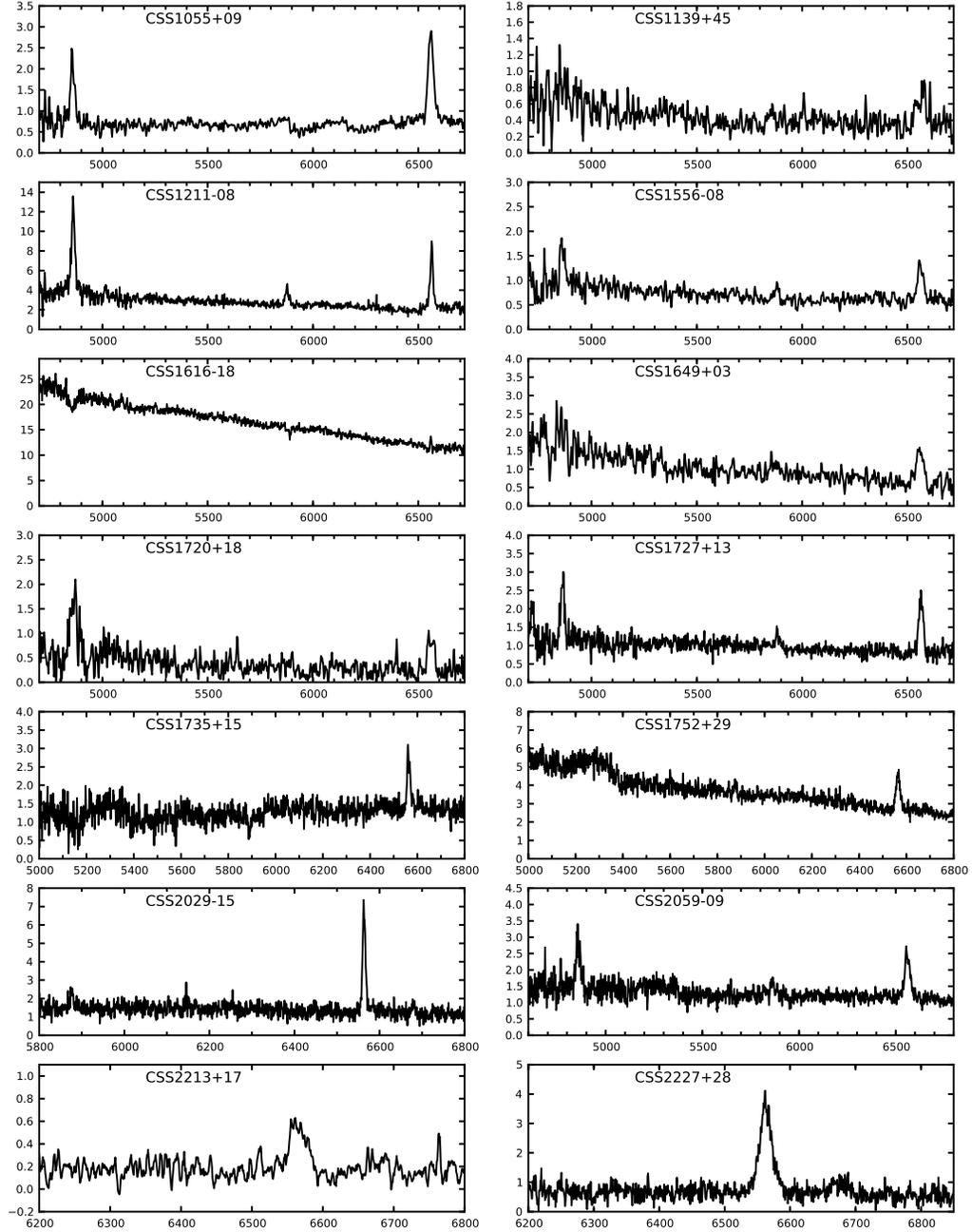}
\caption{Reconnaissance spectra of the remaining 12 CRTS CV candidates.
The scales are the same as in Fig.~\ref{fig:idspecs1}.
}
\label{fig:idspecs2}
\end{figure}

Table \ref{tab:spectro_summary} shows that we
obtained only brief exposures of most of our 
spectroscopic targets, generally taking one or 
two exposures in a single visit.  Our main purpose
was to verify that the objects showed spectra typical of 
CVs.  Figs.~\ref{fig:idspecs1} and \ref{fig:idspecs2} show
these exploratory spectra.  Of the 28 objects surveyed,
27 appear to be bona fide CVs, the exception being 
CSS0350+35.  Here are brief descriptions. 

{\it CSS0051+20.}  Our one spectrum has modest signal-to-noise,
but shows clear H$\alpha$ emission and confirms that this is a
CV.  The continuum has a shape that hints at an 
M-dwarf contribution, but we cannot be sure this is real.

{\it CSS0208+37.}  The Balmer emission lines are strong and
appear double-peaked, suggesting a low orbital inclination.
A prominent blue continuum may be from an accretion disk, or
may be from a white dwarf photosphere; on the other hand, the
Balmer decrement appears extreme, with H$\beta$ much
stronger than H$\alpha$, suggesting that an 
instrumental effect might be enhancing the blue end
of the spectrum.  Dwarf novae declining from outburst
can sometimes show blue continua, but our synthesized and 
measured magnitudes
($V = 18.3$ and $V = 18.23$, respectively) are both 
a bit fainter than the CRTS minimum magnitude of
17.9, ruling out this interpretation; the filter 
photometry was obtained only two days before the 
spectrum. 

{\it CSS0328+28.}  This is an unremarkable dwarf
nova spectrum.

{\it CSS0350+35.} The spectrum is a good match to an
M1 dwarf, and shows only a narrow H$\alpha$ line
similar to a dMe star.   While this might be a 
CV in which mass transfer has stopped, there is 
no sign of a white dwarf in the spectrum, so
the spectrum is consistent with a flare star.
The light curve at the CRTS does not rule out
either a CV or flare star, so the flare star
classification appears likely.
This is, notably, the 
{\it only} object in our spectroscopic sample 
that appears not to be a CV.

{\it CSS0401+08.} Our spectrum is consistent with a 
dwarf nova in outburst, with weak H$\alpha$
emission and weak, broad H$\beta$ absorption 
(with a hint of a central reversal) on a blue 
continuum.  The magnitude synthesized from the 
spectrum ($V = 16.5$) is also well above the 
18.5 mag minimum listed in the CRTS.

{\it CSS0411+23.} The emission lines are relatively
narrow (H$\alpha$ has a FWHM near 300 km s$^{-1}$),
suggesting a relatively low orbital inclination.  
Weak helium emission is present at $\lambda \lambda 
6678, 5876, 4921$, and 5015.  

{\it CSS0440+02.}  The emission lines are broad
($\sim 1900$ km s$^{-1}$ FWHM) and just show 
double peaks at our signal to noise ratio.  The 
system is probably close to edge-on.  

{\it CSS0447+09.} This shows evidence for a 
K-type contribution, in the form of an 
absorption dip near 5170 \AA , and absorption 
in the Na D lines.  The signal-to-noise
ratio is not good enough to warrant a detailed
decomposition of the spectrum, but the absorption
features suggest that about half the light comes
from a mid-K star, probably in the range
K4 to K6.  This suggests that the 
orbital period is likely to be $\sim 6$ h
or greater \citep{kniggedonor}, but a much shorter
period is possible if the secondary has lost
much of its mass \citep{thorstenseneiuma}.

{\it CSS0505+08.}  We detect H$\alpha$ at 
rather low signal-to-noise ratio, confirming
the CV nature of the object.  

{\it CSS0514+08.} Both H$\alpha$ and 
H$\beta$ are detected in emission.  The 
FWHM of H$\alpha$ is around 1000 km s$^{-1}$,
suggesting an intermediate orbital inclination.
No secondary star is seen at our signal-to-noise
ratio. 

{\it CSS0518$-$02.} This object shows an 
unusual spectrum, with a blue continuum and 
a narrow ($\sim 300$ km s$^{-1}$), relatively
weak H$\alpha$ emission line.  The NaD lines
are detected in absorption, but the continuum
does not show any other convincing late-type 
features, so the NaD may be interstellar.  
The spectrum's synthetic magnitude 
($V = 16.9$) suggests that the object was somewhat 
brighter than minimum light ($m = 17.5$), but
not dramatically so.  The CRTS light curve shows
many outbursts of $\sim 2$ magnitudes.  This may
be a Z Cam-type dwarf nova, with a persistent
plateau state near $m = 17.5$, or could be
some other kind of novalike variable.

{\it CSS0545+02.} The red Digital Sky Survey
and the SDSS finding chart (linked from the 
CRTS site) show a nebulosity around this object,
strongest along a northeast-southwest axis,
extending to a radius of about 20 arcsec from
the object.  In our spectrum, the apparent
sharp absorption features at H$\alpha$ 
(6563 \AA), [NII] ($\lambda\lambda$ 6548 and
6584), and [SII] ($\lambda\lambda$ 6717 and
6730) are strong nebular emission features in the sky
that have been over-subtracted.  The long-slit
spectrum shows extended weak H$\alpha$, H$\beta$,
[NII] and [SII] emission along 2.6-arcmin
length of the slit, and [OIII] $\lambda\lambda$ 4959 and
5007 emission extending to $\sim 18$ arcsec 
from the star.  
The stellar spectrum shows H$\alpha$
emission with a width of $\sim 2000$ km s$^{-1}$, 
very much resembling a CV.  The CRTS light curve
hangs mostly at $m = 16.5$ or so, but the 
SDSS has $g = 18.7$, in good agreement with our
$V = 18.75$ and $V = 18.9$ synthesized from 
our spectrum; the nebula may be affecting the 
CRTS measurement.  If this really is a CV, the 
nebulosity is especially intriguing; it is also
possible that this is a planetary nebula central
star, in which case the broad H$\alpha$ might arise
from a stellar wind.  The CRTS light curves shows some
faint upper limits, so the system may eclipse.

{\it CSS0558+00.}  Our single, low
signal-to-noise spectrum shows weak,
broad H$\alpha$ on a blue continuum.  The
synthesized magnitude ($V = 16.6$) is well above
the minimum $m = 19.0$ found by CRTS.  It appears
we caught this in a relatively rare outburst; 
the CRTS light curve shows only three outbursts, 
even though the object is well-observed.  Our
filter photometry, obtained 9 days earlier than the 
spectrum, shows the object much fainter, at 
$V = 20.45$.  This is a dwarf nova.

{\it CSS0905+12.} This appears to be an 
ordinary dwarf nova observed near minimum
light. H$\alpha$ has a FWHM of 1200 km s$^{-1}$,
suggesting an intermediate orbital inclination.

{\it CSS1055+09.}  The spectrum shows a
contribution from an M dwarf which is 
presumably the secondary star.  Because of 
our limited spectral coverage and signal-to-noise,
we can only roughly estimate the secondary's spectral type 
to be M3 $\pm 2$.  The secondary appears to contribute at
least half the continuum light in the region of 
our spectrum.  The easy visibility of the secondary
suggests that the orbital period lies above the 2-3 hour
gap.  

{\it CSS1139+45.} Our spectrum is very poor, but
shows broad H$\alpha$ emission, confirming that this
is a CV.  The synthesized magnitude and spectrum
both indicate it was at minimum light.

{\it CSS1211$-$08.} This relatively bright 
object shows strong Balmer and HeI lines.  
H$\alpha$ has a FWHM of 800 km s$^{-1}$, indicating
a fairly low orbital inclination, and the spectrum
and synthesized magnitude indicate that the spectrum 
was taken near minimum light.

{\it CSS1556$-$08.} Our spectrum has poor signal-to-noise,
but does show broad H$\alpha$ and H$\beta$ emission.  
The synthesized
$V = 19.3$ is fainter than the CRTS minimum, but 
conditions were partly cloudy at the time.

{\it CSS1616$-$18.}  The spectrum is typical
of dwarf novae in outburst, with narrow H$\alpha$
emission, broad H$\beta$ absorption, and a blue
continuum.  The synthesized $V = 15.9$, while the
CRTS lists variation between $m = 14.9$ and $m = 17.8$.

{\it CSS1649+03.} Broad H$\alpha$ emission is 
clearly detected, but the poor signal-to-noise ratio 
precludes further analysis.  Our photometric measurements
and the synthesized magnitudes both agree well with the
magnitude at minimum listed in the CRTS.

{\it CSS1720+18.}  Strong, broad H$\alpha$ and H$\beta$
confirm the CV status.  Our spectrum has 
poor signal-to-noise, which is unsurprising given the
object's faintness.

{\it CSS1727+13.} H$\alpha$ has a FWHM of 1200
km s$^{-1}$, indicating an intermediate inclination.
The spectrum is typical of dwarf novae at minimum
light, and has a synthesized $V = 19.0$.  It was taken
only 3 days after photometric measurements showing 
$V = 16.96$, indicating that the star had
faded quickly following an outburst.

{\it CSS1735+15.} Our spectrum was taken in partly
cloudy conditions, but does show H$\alpha$ in 
emission.  In addition, NaD absorption is 
present, and the K-star features near 
$\lambda 5168$ are just detected.  The orbital
period is probably well longward of the 2-3
hr gap.  We think that the hump in the continuum 
near 5300 \AA\ is probably an artefact.

{\it CSS1752+29.} The H$\alpha$ emission has a 
modest equivalent width, and the continuum is
blue, suggesting that the system was in outburst.
Magnitudes from the OSMOS acquisition image and the 
synthesized spectrum agree nicely at $V \sim 17.3$, 
while the CRTS lists $m = 18.3$ for minimum light, so
the system was likely declining from outburst.  
Again, the continuum hump near 5300 \AA\ is probably
an artefact. 

{\it CSS2029$-$15 = SY Cap.}  \citet{kato09}  
found superhumps 
with $P_{\rm sh}$ = 0.063759(22) d = 91.8 min.
Using the $P_{\rm orb}$-$P_{\rm sh}$ relation derived
by \citet{unveils}, we predict $P_{\rm orb} = 89.8$ min. 
Our spectrum shows narrow H$\alpha$  (FWHM $\sim$
400 km s$^{-1}$), indicating a low orbital inclination.

{\it CSS2059$-$09.} The H$\alpha$ line has a 
FWHM of 1500 km s$^{-1}$, suggesting a fairly low
orbital inclination.  
The CRTS light curve shows a
gradual brightening of the minimum magnitude over
time, with outbursts superposed.  Our spectrum and
acquisition image were taken in partly cloudy conditions.

{\it CSS2213+17.} The H$\alpha$ portion of our 
spectrum was unaffected by a dewar condensation 
problem, and shows a broad emission line, confirming
that this is a CV.  

{\it CS2227+28.} This spectrum was also affected by 
the condensation, but H$\alpha$ was again in the 
clear portion, and is nicely detected with a somewhat
triangular profile, indicating a moderate orbital
inclination.

\section{Radial Velocity Studies}
\label{sec:rvs}

\begin{figure}
\plotone{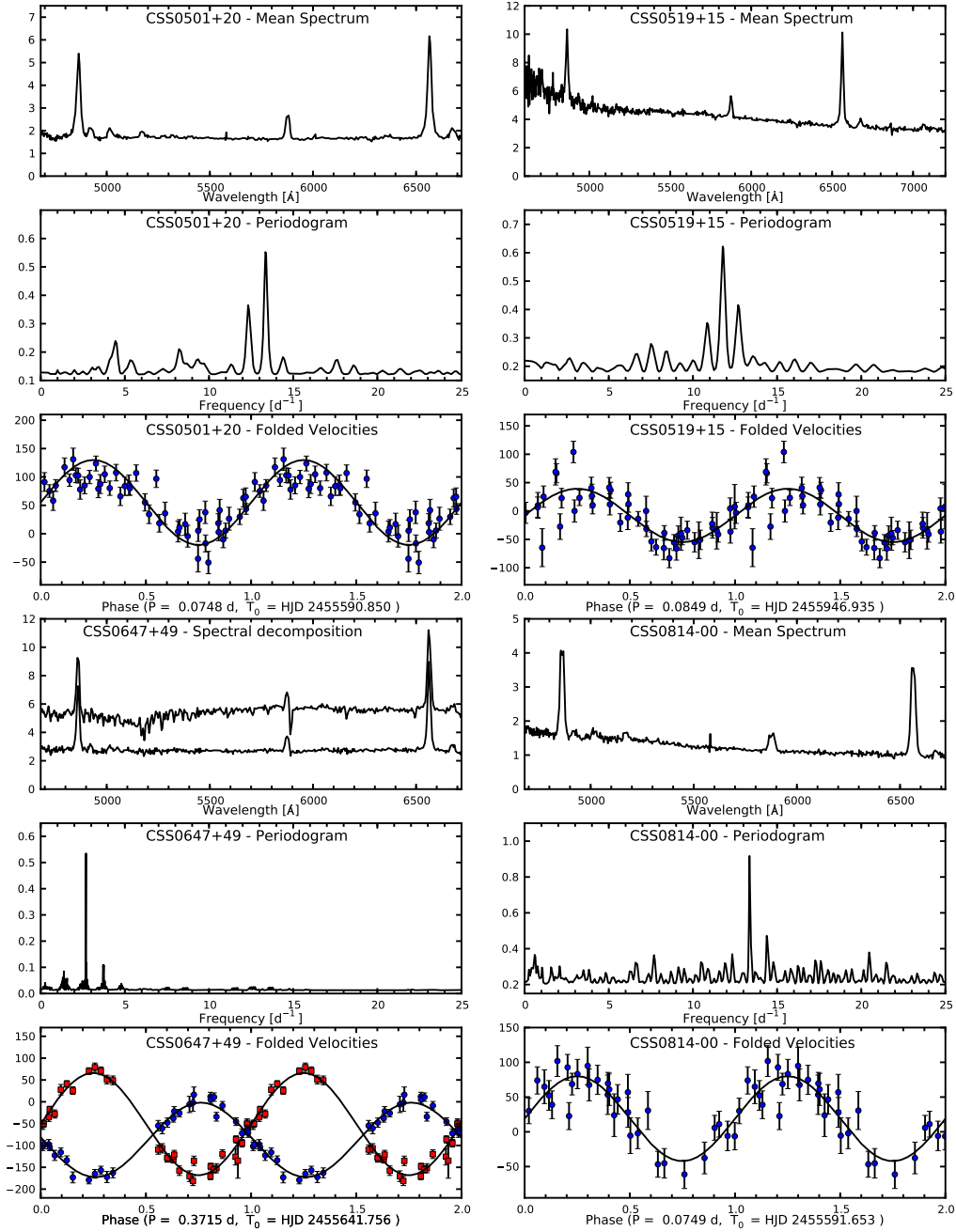}
\caption{Mean spectra, periodograms, and folded velocity curves for four of the 
objects studied here.  The vertical axes for the spectra are
in units of $10^{-16}$ erg cm$^{-2}$ s$^{-1}$ \AA$^{-1}$; for the 
periodograms, the axis is unitless ($1 / \chi^2$); and for the 
radial velocities, the axis is in km s$^{-1}$.  
In the velocity curves, all data are repeated on an extra cycle for 
continuity, the uncertainties shown are estimated from 
counting statistics, and the solid curves show the 
best-fitting sinusoids.  For CSS0647+49, both emission and
absorption velocities are plotted, and the lower trace in the 
spectrum plot shows the result of subtracting a scaled spectrum of
a late-type star chosen to match the secondary's contribution (see text).
}
\label{fig:montage1}
\end{figure}

\begin{figure}
\plotone{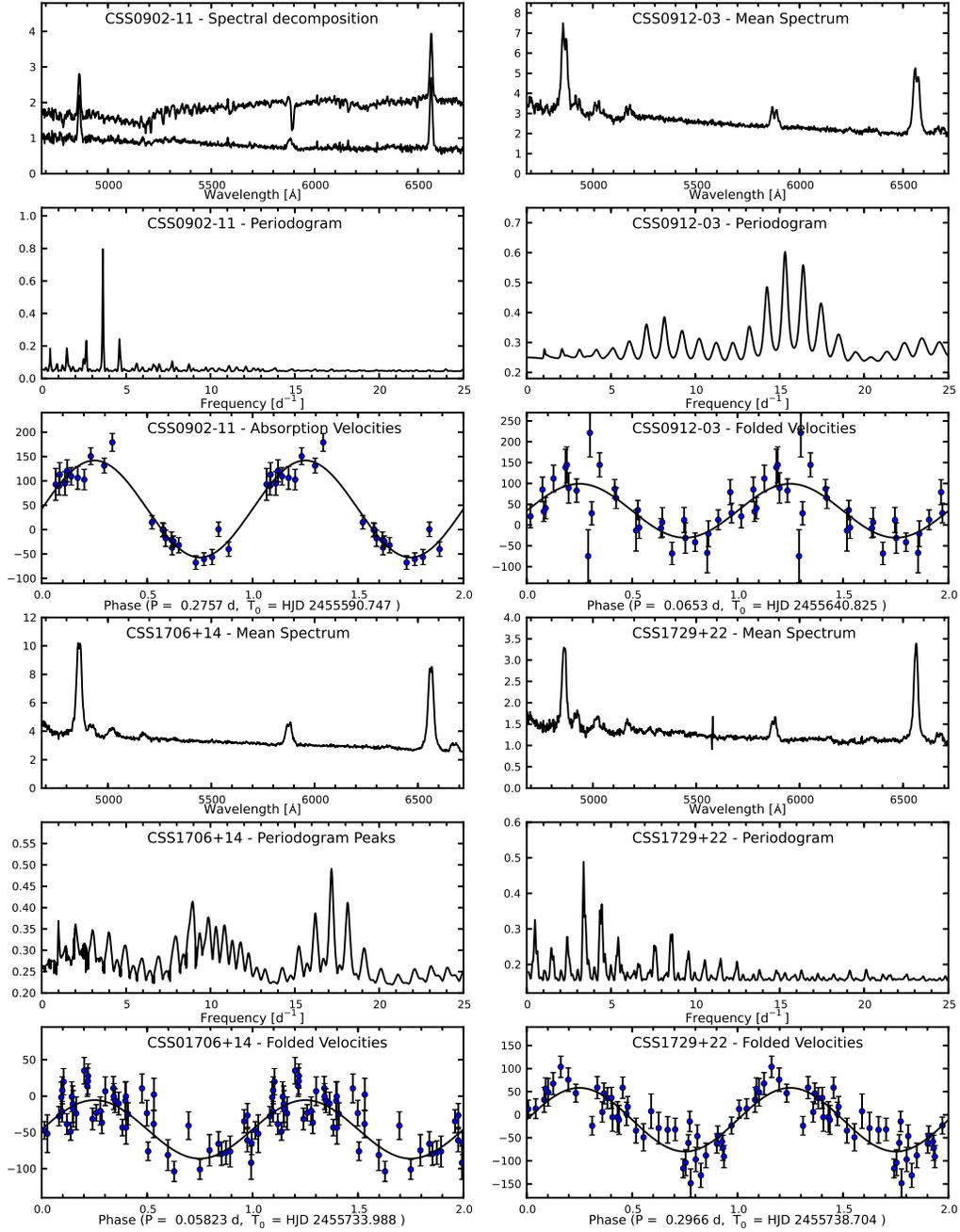}
\caption{
As for Fig.~\ref{fig:montage1}, for the four remaining objects
with radial velocity studies.  For the lower spectrum trace for
CSS0902$-$11, we have subtracted a matched, scaled late-type star. 
}
\label{fig:montage2}
\end{figure}

We obtained time series spectroscopy for eight
targets.  Figs.~\ref{fig:montage1} and \ref{fig:montage2}
show the average spectra, periodograms, and folded
velocity curves, and Table~\ref{tab:parameters}
gives the parameters of the sinusoidal fits.
We discuss the objects in order of RA.

\subsection{CSS0501+20}

The spectrum is typical of dwarf
novae at minimum light.  The lines are
single-peaked, and  H$\alpha$ has a 
FHWM of 1100 km s$^{-1}$.  We adopt
$P_{\rm orb} = 107.7 \pm 0.2$ min. An
alternate choice of daily cycle count
gives 116.7 min, but the Monte Carlo test
described by \citet{tf85} assigns this
a probability below 1 per cent.  At this 
orbital period, this is likely to be an
SU UMa star showing superhumps and superoutbursts.

\subsection{CSS0519+15}

The equivalent widths of the emission lines are
rather smaller than in most dwarf novae at minimum
light, suggesting that we caught the system somewhat
above minimum.  The lines are single-peaked and
the FWHM of H$\alpha$ is $\sim 800$ km s$^{-1}$, 
indicating a fairly low inclination.  Our 
$P_{\rm orb}, 122.3 (4)$ min, places this in the
period range of the SU UMa stars.  

\subsection{CSS0647+49}

The spectrum shows a conspicuous contribution
from a late-type star.  Using spectra of 
stars classified by \citet{keenan89}, we 
find that the companion's spectral type to 
be K4.5 $\pm$ 1 subclass, and that its contribution
to the spectrum is equivalent to $V = 18.0 \pm 0.4$.  
In Fig.~\ref{fig:montage1}, the upper spectral
trace shows the average spectrum, and the lower
shows the result of subtracting a scaled K4 star from 
the average.  
Most of our observations are from
2011 March, but we also have velocities from January
and September.  An unambiguous cycle count over the 
whole interval yields $P_{\rm orb} = 8.9160 \pm 0.0005$
hr.   The emission- and absorption-line velocities 
are both modulated; the emission-line modulation
is $0.510 \pm 0.007$ cycles out of phase from the 
absorption, consistent with the half-cycle offset
expected if the emission lines trace the white dwarf
motion.  If we {\it assume} that this is the case, then
the mass ratio (secondary to white dwarf) is 
$0.73 \pm 0.04$.  With this mass ratio and a
typical white dwarf mass of $M_{\rm wd} = 0.8$ M$_\odot$,
the orbital inclination $i$ would be near 35$^\circ$.  
While this is only meant to be illustrative, the 
rather low secondary velocity amplitude ($K_2 = 117 \pm 4$ 
km s$^{-1}$) implies an orbital inclination 
low enough that eclipses are very unlikely.

We estimate a distance using the secondary's 
contribution as follows.  If we fix $P_{\rm orb}$ at its
measured value and assume that the secondary
star fills its Roche critical lobe, 
then the secondary's radius, 
$R_2$ depends almost entirely on its mass $M_2$; 
furthermore, the dependence of $R_2$ on $M_2$ is weak.  
Evolutionary calculations by \citet{bk00} suggest that a 
K4 star in an 9-hour orbit should have 
$M_2 = 0.7 \pm 0.2$ M$_\odot$.  As a check,
taking our mass ratio at face value and assuming
a broadly typical $M_1 = 0.8$ M$_\odot$ for the
white dwarf gives $M_2 = 0.58$ M$_\odot$, in reasonable
agreement.   With this mass range, the approximation
given in eqn.~1 of 
\citet{beuermann2ndry} constrains 
$R_2$ to be $0.85 \pm 0.10$ R$_\odot$.  
From data in \citet{beuermann99}, we estimate the 
surface brightness of a K4.5 dwarf to be such that
it would have $M_V = 6.3 \pm 0.5$ if it had
$R = 1$ R$_\odot$.  Scaling this to the secondary's
radius yields $M_V = 6.6 \pm 0.6$.  The secondary's
synthetic $V$ magnitude is 18.0, but this is probably 
a little too faint, because the synthetic magnitude
of the average spectrum ($V = 17.2$) is 
fainter than the $V = 16.84$ we find from 
the more accurate filter photometry.  Discrepancies 
this large are to be expected (Section 
\ref{sec:techniques}).
Correcting for these losses gives 
$V \sim 17.6$ for the secondary contribution. 
At this celestial location
($l = 166.19^\circ, b = 19.79^\circ$),
\citet{schlegel98} estimate the total Galactic
extinction to be $E(B-V) = 0.11$, which taking
$R = 3.3$ makes $A_V = 0.36$, assuming the star
lies outside the dust layer.  Putting all this 
together yields $(m - M)_0 = 10.9 \pm 0.7$,
or a distance $d = 1300 (+500,-400)$ pc.
Notice that we did not assume that the secondary
star follows a main-sequence mass-radius relation,
but rather combined the Roche size constraint with the
surface brightness.

\subsection{CSS0814$-$00}

This is an SU UMa-type dwarf nova, and was observed in superoutburst 
by \citet{kato09}, who detected superhumps.  The superhump
period $P_{\rm sh}$ was not determined cleanly, but 
appeared to be  0.0763 d.  Photometry obtained
by \citet{ww10} gave $P_{\rm orb} = 1.796$ h, or
0.0748 d, in rough accordance with expectations based
on the superhump period.  Our spectroscopic period is
essentially identical, at 0.07485(5) d.  The emission 
lines are barely double-peaked, suggesting that the 
inclination is not small, but \citet{ww10} make no 
mention of an eclipse.  As far as we know, this is 
the only system studied here which has a period
determination in the literature.

\subsection{CSS0902$-$11}

Like CSS0647+49, this object also has a strong contribution
from the secondary star.  By comparing the spectrum 
with stars classified by \citet{keenan89},
we estimate the secondary's spectral type to be K7 $\pm$ 1 
subtype, and find that the secondary's contribution is 
nominally equivalent to $V = 19.1$.   However, many of the
exposures used in the average spectrum were taken in 
partly cloudy conditions and mediocre seeing.  
Our average spectrum has synthesized $V = 18.4$, but 
our best exposures have $V = 17.8$.   The CRTS light 
curve shows that the source is fairly steady when not 
in outburst, at
$m = 17.5$, consistent with our best exposures, so
it appears that the mean spectrum from which the secondary
magnitude was derived is about 0.6 mag too faint.
We therefore adopt $V = 19.1 - 0.6 = 18.5$ for the secondary
contribution.

The emission-line radial velocities
did not yield a period, but the absorption spectrum
showed an unambiguous modulation at $6.62 \pm 0.01$ hr.  
The relatively small velocity amplitude of the secondary 
($K_2 = 100 \pm 6$ km s$^{-1}$) constrains the inclination
to be fairly low for any realistic white dwarf mass, so
eclipses are not expected.  

We can once again estimate a distance using the secondary's
contribution, following the same procedure we used
for CSS0647+49.  Using \citet{bk00} as a guide, we estimate 
$M_2 = 0.55 \pm 0.15$ M$_\odot$, from which we infer
$R_2 = 0.68 \pm 0.08$ R$_\odot$ at this period.  
The surface brightness for K7 $\pm$ 1 is equivalent to
$M_V = 7.25 \pm 0.5$ for a 1 R$_\odot$ star \citep{beuermann2ndry}. 
At this location ($l = 239.86^\circ,
b = 22.33^\circ$) the \citet{schlegel98} extinction map
gives $E(B-V) = 0.05$ mag.  Combining this with the 
apparent magnitude of the secondary gives 
$d = 1100 (+350, -260)$ pc. 

\subsection{CSS0912$-$03} 

The emission lines are notably broad -- the FWHM of H$\alpha$
is nearly 1700 km s$^{-1}$ -- and the lines show
incipient double peaks, so the inclination is likely to 
be fairly high.  Even so, the radial velocity
amplitude is modest.  We detect a modulation in the
emission line velocity, but with the  available 
the choice of daily cycle count is ambiguous, so 
we give two possible periods in Table \ref{tab:parameters}.
Both possible periods are well below the 2-3 hour gap, so 
it is likely that this will prove to be an SU UMa-type
dwarf nova with superoutbursts and superhumps.

\subsection{CSS1706+14}

The spectrum is typical of dwarf novae at minimum 
light.  We obtained radial velocities on a single 
night in 2011 June, after which the star went into 
outburst, suppressing the H$\alpha$ emission and 
ending the measurements.  In 2012 May we
obtained velocities on three consecutive nights.
From the combined 
data we adopt a period near 0.0582 d, but
periods near 0.111 d and 0.0552 d (the latter being
a daily cycle count alias of our adopted period) are 
not completely ruled out.  The $\sim$ 345 d gap in the 
time series created fine-scale ringing in the 
periodogram.  To derive the period uncertainty in Table 
\ref{tab:parameters}, we shifted the 2012 May
data back in time by an integer number of periods,
removing the gap.  The exact period used in this 
artificial shift has essentially no effect on the 
resulting period uncertainty.  

\subsection{CSS1729+22} 

The spectrum shows weak M-dwarf features, in particular
the extra continuum around 5950-6170 \AA , and the
band head near $\lambda 6180$.  The features are too
weak to derive good constraints on the spectral type,
but an M1.5 dwarf contributing around 25 per cent of the 
flux at 6500 \AA\ gives a reasonably good match.  As
one might expect from the spectrum, the period is 
relatively long; the best-fitting $P_{\rm orb}$ is
7.12(3) hr, or 3.37 cycle d$^{-1}$; however, we cannot 
entirely rule out an alias at 4.45 cycle d$^{-1}$, or 5.40 hr.

\section{The CRTS Sample and the Cataclysmic Variable Population}
\label{sec:statistics}

As noted earlier, the CRTS CV sample is 
of great interest in characterizing the CV population.
In this section, we consider what the CRTS sample can
tell us about the completeness of the available 
CV sample.

The number of non-CVs included in the CRTS sample 
appears to be small.  In the sample of 36 objects
for which we obtained spectra, we found only a single 
apparent non-CV.  The fainter objects in the CRTS
sample were beyond our magnitude limit, 
and it is possible that the fainter end of the sample
includes a greater fraction of interlopers. 
However, the selection criterion 
-- outbursts of more than 2 mag --
should be fairly robust even for faint objects.
We assume, then,
that essentially all CRTS CVs are real CVs.

\subsection{Other Samples Used for Comparison}

We compared the CRTS list to several other samples of 
CVs, which we describe here.

{\it The SDSS Sample.} SzkodyI-VIII list 286 CVs.  
Although the SDSS Data Release 8 covers some 
areas at low galactic latitude that CRTS does not \citep{sdssviii},
all of the SDSS CVs
lie within the nominal sky coverage of the CRTS, so
for practical purposes the SDSS coverage is entirely
contained within the CRTS coverage.  SzkodyI-VIII 
do not tabulate the subtypes of their objects, though
they do give limited information on this.  To enable
more detailed comparisons, we classified the objects
in SzkodyI-VIII, primarily on the basis of their
spectra, supplemented by the information in the
text of the papers.  Our classification scheme was
as follows:

\begin{description}

\item[DN]  This class was used for objects known to be
dwarf novae, and for objects whose spectra resemble those
of dwarf novae.
The spectra classified this way tended to have strong, 
broad Balmer emission 
(with H$\alpha$ equivalent width usually greater than 30 \AA\ ), 
relatively flat disk continua (in $F_\lambda$ vs.~$\lambda$), and
weak or absent HeII $\lambda$4686.
Objects showing blue continua and white-dwarf
absorption wings around H$\beta$ were classified as
DN-W; if a K or M-type secondary was present, we
classified the object as DN-2.  In some cases the SDSS 
spectrum shows the object in outburst.  Dwarf novae in
outburst can be difficult to distinguish from novalike 
variables, but in these cases the flux level in the 
spectrum will usually be much greater than expected
based on the imaging data.  

\item[NL] This class included spectra showing blue 
continua, without white-dwarf absorption,  and relatively 
weak emission lines, or
stronger emission lines and substantial HeII $\lambda$4686
(typically half the strength of H$\beta$ in those cases).
The Balmer absorption wings in a novalike variable can
superficially resemble white-dwarf absorption, but with
experience the distinctive white-dwarf line profile can
usually be distinguished from the disk absorption lines
seen in UX-UMa type variables.

\item[AM] These showed HeII $\lambda$4686 similar in 
strength to H$\beta$, or other evidence of a
strong magnetic field such as cyclotron humps.

\item[NCV] We assigned 14 objects to this ``Non-CV" class. 
This heterogeneous group include objects whose spectra resembled 
reflection-effect white dwarf systems, subdwarf B stars, and 
chromospherically active M dwarfs.  One, 
SDSS J1023+00, has proven to be a binary 
containing a millisecond radio pulsar \citep{archibald,wang,
thorj1023}.  

\end{description}

Removing 14 apparent non-CVs from the 
SDSS sample leaves us with 272 objects.  
Because of the limited information --
especially the lack of long-term light curves -- the 
classifications should be considered somewhat rough.
They are given in Table \ref{tab:sdssclasses}.

{\it The Ritter-Kolb Catalog.} \citet{ritterkolb} 
have maintained a catalog (hereafter RKcat) of 
CVs and related object with known or suspected orbital
periods.  Some of these were discovered in the CRTS,
but entries that do not match with CRTS objects 
clearly were not.  In our comparisons, we 
used the cataclysmic 
binary list, `cbcat', from version 7.16 of the 
Ritter-Kolb catalog (hereafter RKcat)\footnote{available at 
http://www.mpa-garching.mpg.de/RKcat/}, which 
contains 926 objects, of which 582 are in the
CRTS survey area.
RKcat provides subclassifications similar to the ones
we invented for the SDSS sample.

{\it The General Catalog of Variable Stars.}
We downloaded the General Catalog of Variable Stars 
\citep{gcvs}; the 2012 January version contains 45678
entries, of which 753 are classified as some kind of 
CV (types N, NA, NB, NC, NL, UG, UGSS, UGSU, or AM).
Of these, 287 are in the CRTS survey area.

\subsection{Comparison with Other Samples.}

Table \ref{tab:samplestats}
shows the number of cross-matches, and non-matches
between the CRTS sample and the lists detailed in the
previous section.   

\begin{figure}
\epsscale{0.95}
\plotone{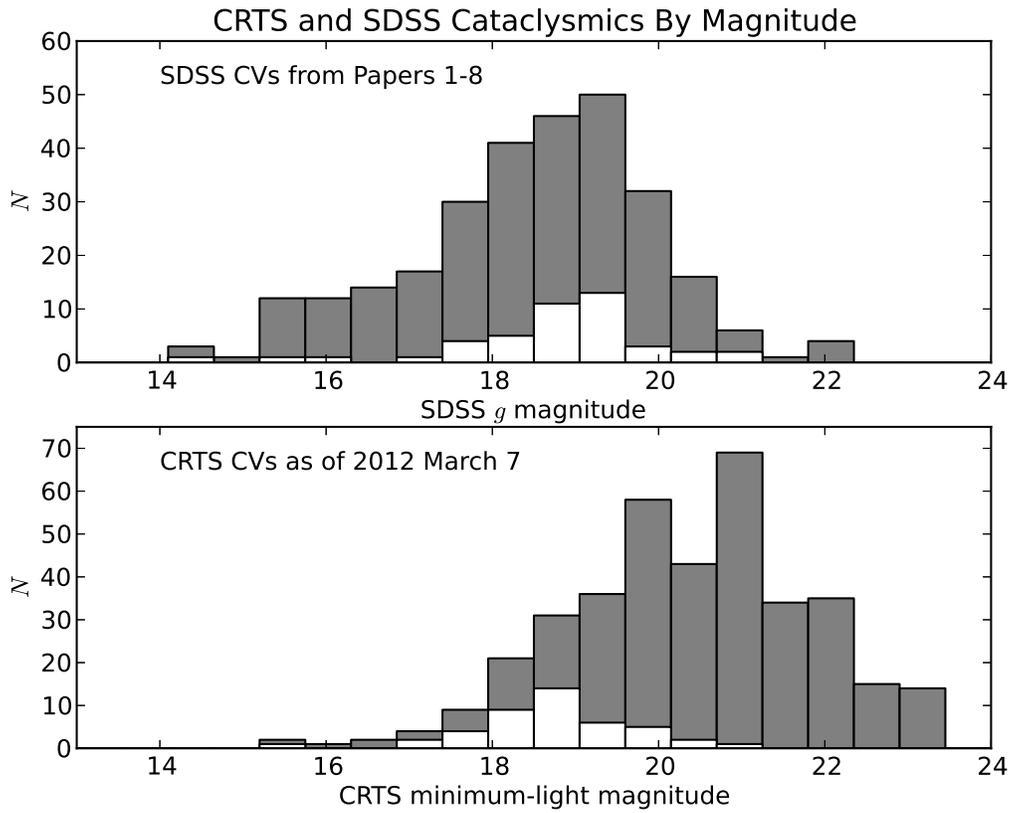}
\caption{Histograms of the SDSS and CRTS samples (see text for 
details).  The gray bars give the full SDSS and CRTS samples, while
the white portions are the subset that are in both samples.  
}
\label{fig:sdsscrtshist}
\end{figure}

There are only 44 cross-matches between the 
CRTS and the SDSS CV samples.  
Fig.~\ref{fig:sdsscrtshist} shows histograms of the 
SDSS sample and the CRTS objects that lie in the 
SDSS footprint.  Note the following:

\begin{enumerate}

\item 
At minimum light, most of 
the CRTS objects are too faint to have been identified
as CVs by SDSS.  The fainter CRTS objects are for the most part
detected in the SDSS imaging data (provided they were
in the SDSS footprint),
but were too faint to be selected for spectra.

\item Among the brighter CRTS objects, somewhat more than 
half are not in the SDSS sample. 

\item The top panel shows that 
only a rather small fraction of SDSS objects are recovered
by CRTS.  

\end{enumerate}

Why does CRTS miss so many CVs?  As
noted earlier, one expects CRTS to 
preferentially select dwarf novae, and to
be less sensitive to other CV subtypes.
Consistent with this, 40 of the 44
cross-matches between SDSS and CRTS are objects 
that we classified as `DN'.  The RKcat and
GCVS comparisons (Table \ref{tab:samplestats})
also show this tendency for CRTS to select dwarf novae;
of the 134 CRTS objects that are listed in SDSS,
RKcat, or GCVS, 123 of them are classified as 
dwarf novae in at least one of these catalogs, and only 
11 are other kinds of CV.  We therefore confirm that,
as one might expect, CVs that are not dwarf novae
are largely passed over by CRTS.

The objects recovered by CRTS are mostly dwarf novae,
but are the objects {\it not} recovered by CRTS 
{\it not} dwarf novae?
In the latter part of the table, we give the numbers 
of CVs that lie in the nominal CRTS survey area, but which 
are {\it not} recovered by CRTS. 
As expected, dwarf novae constitute a smaller portion of 
the unrecovered objects; the aggregate dwarf nova fraction 
is 348/602, rather than 123/134. 

However, these figures also imply that over half the 
{\it unrecovered} objects actually {\it are}
dwarf novae.  Somehow, 348 dwarf novae that we know of
in the CRTS survey area have slipped through its net.
How can we account for this?

Some of these non-recoveries (or `misses') 
are to be expected, because dwarf novae from
one subclass -- the WZ Sge stars -- erupt very infrequently, 
in some cases on timescales of decades or more.  An
even more extreme example, the star
GD 552 \citep{undana-gd552}, appears identical to a
short-period dwarf nova at minimum light, but it has 
{\it no} observed outbursts.  Some fraction of the 
WZ Sge stars will have been missed simply because they
have not outburst during the $\sim 5$ years that
CRTS has operated. 

To explore the effect of outburst interval on CRTS selection,
we used data from RKcat, which
gives an average outburst interval (which they denote $T1$)
for some of the listed dwarf novae. 
We arbitrarily chose 700 days, roughly half the time span
of the CRTS survey, as the dividing line between
short and long outburst intervals.  In addition, we assigned to the
long-interval group any 
dwarf nova subclassified as `WZ'.
Many dwarf novae could not be assigned because 
because $T1$ was not given (and they 
were not classed as WZ), but sufficient information 
existed to classify 137 dwarf novae from 
the CRTS footprint; 95 of these were in the 
short-interval group, and 42 in the long-interval group. 
The objects that {\it were}
recovered in CRTS included 18 from the short-interval group, and
9 from the long-interval group;  
among RK dwarf novae that were {\it not}
recovered by CRTS, there were 77 short-interval systems, and 33 long-interval
systems.  The fact that so many short-interval systems are
{\it not} recovered by CRTS argues strongly that long outburst intervals
are not the main reason CRTS is missing so many dwarf novae (although
it must account for some cases).  Indeed, the ratio of
short to long interval dwarf novae for CRTS/RK matches is
remarkably similar to the ratio in the group of RK dwarf novae 
that are not matched to CRTS.

\begin{figure}
\epsscale{0.95}
\plotone{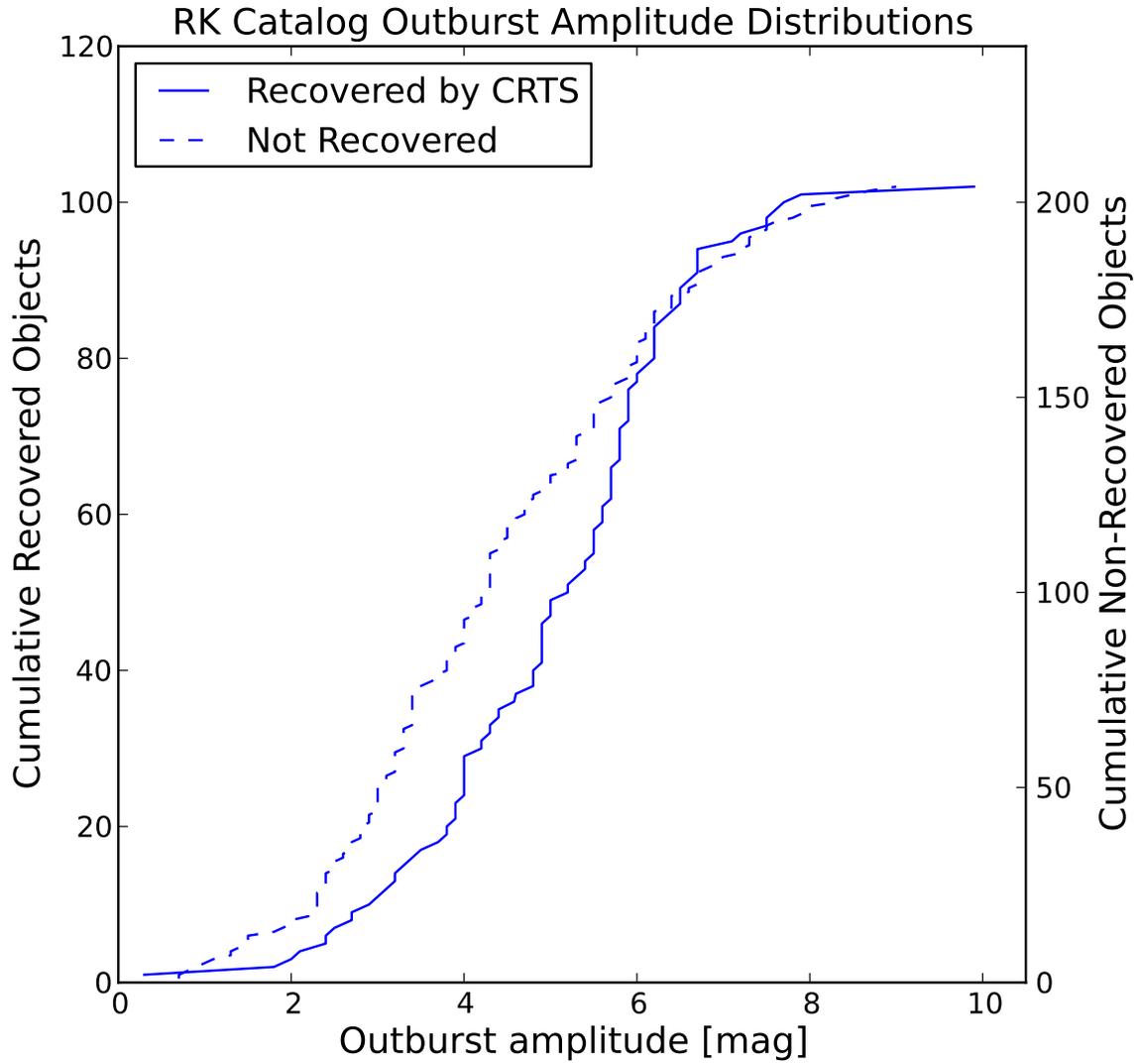}
\caption{Solid curve: Cumulative distribution functions of the outburst amplitudes
of objects in RKcat that are recovered in the CRTS.
Dashed curve:  The same, for RKcat objects that are in the
CRTS footprint, but were not recovered by CRTS.
}
\label{fig:outbcdf}
\end{figure}

The main reason for the incompleteness must therefore lie elsewhere.  
Dwarf nova outbursts will be missed if they occur between observations.
Perusal of CRTS light curves suggests that some parts of the 
sky are covered rather infrequently.
A good number of dwarf nova outbursts must
therefore `slip through the cracks' in the coverage.  This
is exacerbated by the 2-magnitude criterion; not only must
the object be caught in outburst, but during that part of the 
outburst where it is more than 2 magnitudes above minimum.
To see whether outburst amplitude might be an important
selection factor, we again turned to RKcat, this time
finding the outburst amplitude $\Delta m$.  For objects
that did not have superoutburst magnitudes listed, we 
let $\Delta m = mag1 - mag3$ (in their notation), but
if superoutburst magnitudes were available, we used
$mag1 - mag4$.  Fig.~\ref{fig:outbcdf} shows the cumulative 
distribution functions of $\Delta m$ for the recovered and 
unrecovered dwarf novae.  As expected, CRTS is
nearly blind to objects with $\Delta m < 2$; more
interestingly, there is a significant bias against
smaller outburst amplitudes extending all the way 
up to $\Delta m = 6$.  This effect probably arises
because, in any given snapshot, a large-amplitude object
will have a greater likelihood of being caught at
$\Delta m > 2$ than a small-amplitude object, which 
would have to be fairly close to its peak brightness
in order to exceed the survey threshold. 

\begin{figure}
\epsscale{0.95}
\plotone{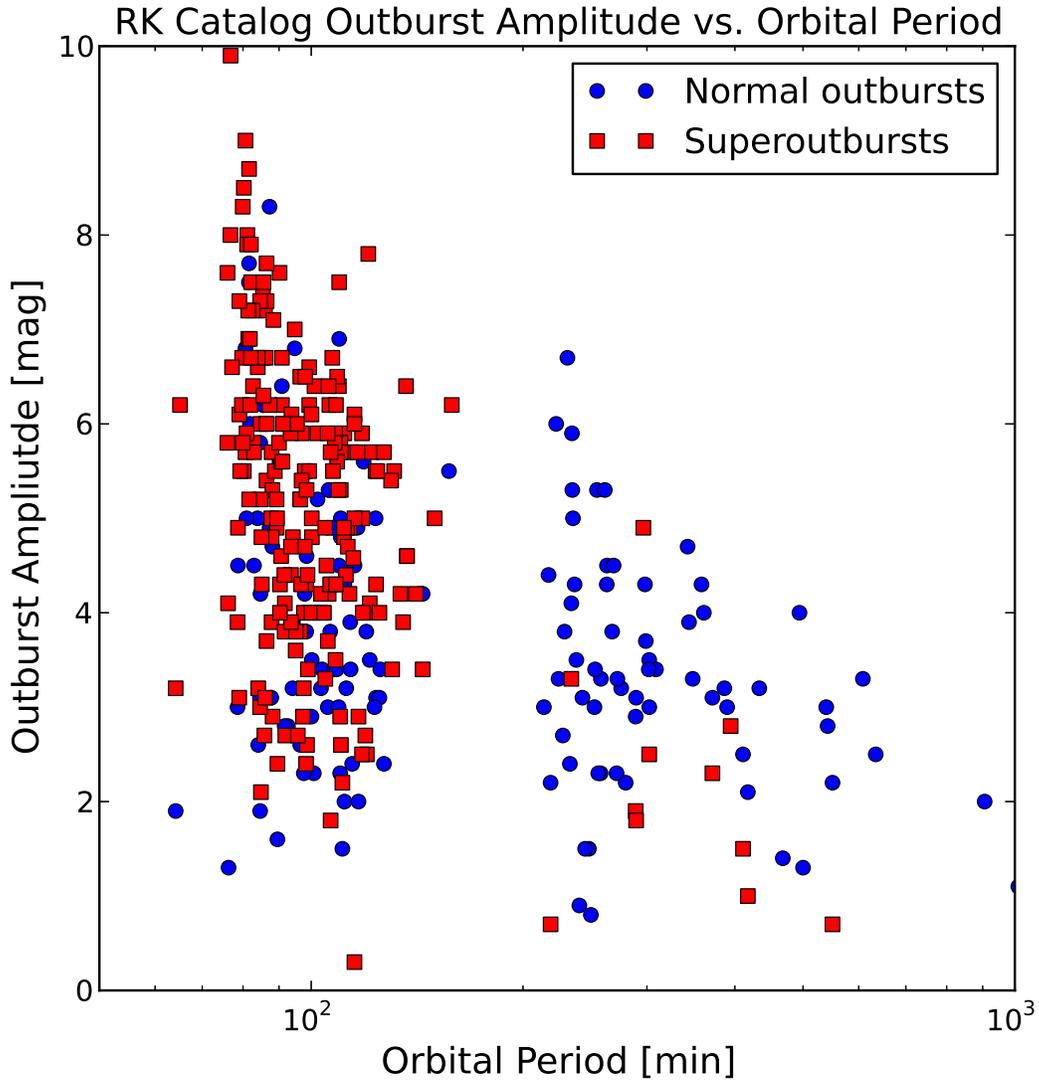}
\caption{Outburst amplitude $\Delta m$ plotted against orbital
period $P_{\rm orb}$, for RKcat objects that have both 
tabulated. 
}
\label{fig:amplvsper}
\end{figure}

It seems, then, that the CRTS survey is biased toward
large outburst amplitudes.  How might this affect other
quantities?  In Fig.~\ref{fig:amplvsper} 
we plot $\Delta m$ against $P_{\rm orb}$, 
for those dwarf novae in the CRTS footprint that have
both quantities tabulated in RKcat.  While there
is a great deal of scatter at any given $P_{\rm orb}$,
there is a clear trend for short-period dwarf novae
to have greater outburst amplitude, and hence a greater
likelihood of being discovered by CRTS.  The CRTS sample
should therefore be biased toward shorter orbital periods.
This must contribute at some level to the preponderance of 
short periods found in this paper and by \citet{ww12}.

The fact that so many of the CRTS CV sample are new discoveries,
and its continuing effectiveness in finding new ones, both show
that a great many CVs remain undiscovered.  Future
synoptic surveys with faster cadence, and
less-stringent variability criteria should discover many
more CVs.

\section{Summary}

We obtained spectra of 36 CRTS CV candidates, and confirmed
that all save one appear to be {\it bona fide} CVs.  For eight 
of the objects we obtained spectroscopic periods, and found that 
three of them had $P_{\rm orb}$ longward of the 2-3 hour gap.
In addition, we examined the overlap between the CRTS CV sample
and others previously-existing samples.  Most CRTS CVs are new
discoveries, but CRTS has not recovered even a majority of the known
dwarf novae in its footprint.   This suggests that a great many
CVs remain undiscovered.  Analysis of the recovered and
unrecovered samples shows that the CRTS sample is biased toward
large outburst amplitudes, which in turn biases it toward
shorter orbital periods.

\acknowledgments
We gratefully acknowledge support from NSF grants 
AST-0708810 and AST-1008217, and thank Dartmouth undergraduates
Jason Spellmire and Erin Dauson for conscientious and
cheerful assistance at the telescope.

\clearpage

\begin{deluxetable}{lrrrrrr}
\tablecolumns{7}
\footnotesize
\tablewidth{0pt}
\tablecaption{List of Observed Objects}
\tablehead{
\colhead{CRTS Name} &
\colhead{Abbreviation} &
\colhead{$N_{\rm outb}$} &
\colhead{$m_{\rm max}$} &
\colhead{$m_{\rm min}$} &
\colhead{Spectra} &
\colhead{Images} \\
\colhead{} &
\colhead{} &
\colhead{} &
\colhead{(mag)} &
\colhead{(mag)} &
\colhead{} &
\colhead{} \\
}
\startdata
CSS091026:005153+204017  & CSS0051+20  &  3  &  15.4 &  18.2 &     2      &   3      \cr 
CSS101009:005825+283304  & CSS0058+28  &  1  &  14.1 &  19.5 &   \nodata  &   3      \cr 
CSS091009:010412$-$031341  & CSS0104$-$03  &  1  &  17.2 &  18.8 &   \nodata  &   3      \cr 
CSS080921:010522+110253  & CSS0105+11  &  4  &  16.4 &  20.0 &   \nodata  &   3      \cr 
CSS091016:010550+190317  & CSS0105+19  &  4  &  16.3 &  19.6 &   \nodata  &   3      \cr 
CSS080922:011307+215250  & CSS0113+21  &  4  &  15.1 &  20.2 &   \nodata  &   3      \cr 
CSS081220:011614+092216  & CSS0116+09  &  3  &  16.1 &  19.2 &   \nodata  &   3      \cr 
CSS091029:015021$-$124425  & CSS0150$-$12  &  6  &  16.3 &  19.1 &   \nodata  &   3      \cr 
CSS101207:020804+373217  & CSS0208+37  &  3  &  15.5 &  17.9 &     2      &   5      \cr 
CSS090928:032812+280631  & CSS0328+28  &  6  &  16.2 &  18.2 &     5      &   4      \cr 
CSS081030:035035+353247  & CSS0350+35  &  9  &  16.5 &  19.2 &     2      &   2      \cr 
CSS081024:040141+080008  & CSS0401+08  &  5  &  15.4 &  18.7 &     2      & \nodata  \cr 
CSS081118:041139+232220  & CSS0411+23  &  3  &  15.5 &  18.3 &     2      & \nodata  \cr 
CSS090219:044027+023301  & CSS0440+02  &  4  &  16.2 &  18.5 &     2      & \nodata  \cr 
CSS081201:044725+092439  & CSS0447+09  &  5  &  15.4 &  18.0 &     3      & \nodata  \cr 
CSS091024:050124+203818  & CSS0501+20  &  9  &  16.4 &  17.8 &    52      & \nodata  \cr 
CSS100321:050527+083415  & CSS0505+08  &  1  &  16.1 &  18.0 &     2      &   1      \cr 
CSS101128:051458+083503  & CSS0514+08  &  4  &  14.3 &  18.5 &     3      & \nodata  \cr 
CSS090219:051815$-$024503  & CSS0518$-$02  & 15  &  15.6 &  17.5 &     2      & \nodata  \cr 
CSS111118:051923+155435  & CSS0519+15  &  8  &  14.5 &  17.7 &    47      & \nodata  \cr 
CSS111003:054558+022106  & CSS0545+02  &  3  &  14.5 &  18.7 &     2      &   5      \cr 
CSS100114:055843+000626  & CSS0558+00  &  3  &  15.7 &  19.0 &     1      &   5      \cr 
CSS091029:064729+495027  & CSS0647+49  &  1  &  13.3 &  16.3 &    33      &   4      \cr 
CSS110413:065037+413053  & CSS0650+41  &  6  &  16.7 &   1.0 &   \nodata  &   5      \cr 
CSS080409:081419$-$005022  & CSS0814$-$00  &  2  &  14.8 &  19.0 &    31      &   5      \cr 
CSS090224:082124+454135  & CSS0821+45  &  6  &  16.8 &  19.8 &   \nodata  &   7      \cr 
CSS090201:090210$-$113032  & CSS0902$-$11  &  5  &  13.0 &  17.5 &    28      & \nodata  \cr 
CSS091022:090516+120451  & CSS0905+12  &  4  &  16.0 &  19.7 &     2      & \nodata  \cr 
CSS110114:091246$-$034916  & CSS0912$-$03  &  7  &  15.3 &  18.0 &    34      & \nodata  \cr 
CSS080130:105550+095621  & CSS1055+09  &  6  &  15.3 &  18.5 &     5      & \nodata  \cr 
CSS081117:113951+455818  & CSS1139+45  &  2  &  15.2 &  19.5 &     2      & \nodata  \cr 
CSS110205:121119$-$083957  & CSS1211$-$08  &  2  &  15.2 &  17.8 &     1      & \nodata  \cr 
CSS100315:121925$-$190024  & CSS1219$-$19  &  0\tablenotemark{a}  &  18.2 &  19.0 &   \nodata  &   8      \cr 
CSS100531:134052+151341  & CSS1340+15  &  1  &  14.5 &  18.0 &   \nodata  &   5      \cr 
\tablebreak
CSS090321:155631$-$080440  & CSS1556$-$08  &  6  &  15.7 &  18.4 &     2      & \nodata  \cr 
CSS100408:161637$-$181027  & CSS1616$-$18  &  8  &  14.9 &  17.8 &     2      & \nodata  \cr 
CSS080131:163943+122414  & CSS1639+12  & 11  &  17.3 &  19.2 &   \nodata  &   4      \cr 
CSS090416:164413+054158  & CSS1644+05  &  6  &  17.7 &  19.8 &   \nodata  &   4      \cr 
CSS100707:164950+035835  & CSS1649+03  &  5  &  14.1 &  18.9 &     1      &   8      \cr 
CSS110426:170152+132131  & CSS1701+13  & 10  &  18.4 &  19.3 &   \nodata  &   4      \cr 
CSS090205:170610+143452  & CSS1706+14  &  2  &  14.8 &  17.9 &    30      & \nodata  \cr 
CSS090929:172039+183802  & CSS1720+18  &  5  &  16.2 &  19.8 &     1      &   4      \cr 
CSS090929:172734+130513  & CSS1727+13  &  7  &  17.3 &  19.5 &     2      &   8      \cr 
CSS100707:172952+220808  & CSS1729+22  &  5  &  15.1 &  17.8 &    46      & \nodata  \cr 
CSS110623:173517+154708  & CSS1735+15  &  1  &  14.2 &  17.3 &     2      & \nodata  \cr 
CSS110610:175253+292219  & CSS1752+29  &  7  &  16.4 &  18.3 &     2      &   1      \cr 
CSS091010:202948$-$155437\tablenotemark{b}  & CSS2029$-$15  &  7  &  14.5 &  18.0 &     3      &   6      \cr 
CSS090528:205933$-$091616  & CSS2059$-$09  & 11  &  15.7 &  18.0 &     2      &  1   \cr 
CSS100404:210954+163052  & CSS2109+16  &  4  &  16.8 &  19.3 &   \nodata  &   4      \cr 
CSS100520:214426+222024  & CSS2144+22  &  2  &  14.7 &  17.1 &   \nodata  &   4      \cr 
CSS090622:215636+193242  & CSS2156+19  &  2  &  17.1 &  18.5 &   \nodata  &   4      \cr 
CSS100615:215815+094709  & CSS2158+09  &  1  &  13.2 &  17.6 &   \nodata  &   4      \cr 
CSS090917:221344+173252  & CSS2213+17  &  2  &  17.2 &  19.0 &     3      &   1      \cr 
CSS090531:222724+284404  & CSS2227+28  &  2  &  14.5 &  18.0 &     2      &   5      \cr 
CSS100520:223234+185036  & CSS2232+18  &  3  &  16.5 &  18.8 &   \nodata  &   4      \cr 
CSS090929:232716+413149  & CSS2327+41  &  3  &  15.9 &  18.4 &   \nodata  &   4      \cr 
\enddata
\tablecomments{
The CRTS name encodes the date of outburst before the colon, and the J2000
celestial coordinates after the colon.
The number of outbursts (column 3) is from a perusal of the light curves (see text).
Magnitudes at maximum and minimum are from CRTS.  Standardized magnitudes and colors
for some of the objects can be found in Table \ref{tab:photometry}.  The last two columns give the
total numbers of spectra and direct images we obtained.
}
\tablenotetext{a}{The CRTS light curve for CSS1219-19 shows a secular increase
from $m \sim 19.5$ to $m \sim 17.5$, with apparently significant 
short-term variability, but no clearly-defined outbursts.}
\tablenotetext{b}{CSS2029$-$15 has been named SY Cap.}
\label{tab:star_info}
\end{deluxetable}

\clearpage

\begin{deluxetable}{llrrrr}
\tablecolumns{7}
\footnotesize
\tablewidth{0pt}
\tablecaption{Summary of Spectroscopy}
\tablehead{
\colhead{Object} &
\colhead{Time} &
\colhead{Instrument} &
\colhead{Exposure} &
\colhead{$V$} &
\colhead{EW(H$\alpha$)} \\
\colhead{} &
\colhead{(JD)} &
\colhead{} &
\colhead{(sec)} &
\colhead{(mag)} &
\colhead{(\AA )} \\ 
}
\startdata
CSS0051+20	&  5944.57  & MN & 600 &  19.0  & 41 \\
CSS0208+37	&  5587.58  & MT & 1080 &  18.3 & 56 \\
CSS0328+28	&  5588.62  & MT &  960 &  19.0 & 48 \\
CSS0350+35	&  5818.93  & OS & 1200 &  18.3 & 1.5\tablenotemark{a} \\
CSS0401+08	&  5587.66  & MT &  960 &  16.5 & 4\tablenotemark{b} \\
CSS0411+23	&  5587.68  & MT & 1080 &  18.9 & 33 \\
CSS0440+02	&  5589.66  & MT & 1800 &  19.3 & 130 \\
CSS0447+09	&  5589.64  & MT & 1800 &  18.9 & 22 \\
CSS0501+20	&  5591     & MT & 27300 &  18.4 & 70 \\
CSS0505+08	&  5820.98  & OS & 1200 &  18.5 & 23 \\
CSS0514+08	&  5588.68  & MT & 1800 &  20.0 & 58 \\
CSS0518$-$02	&  5588.70  & MT &  960 &  16.9 & 4\tablenotemark{b} \\
CSS0519+15	&  5947     & MN & 27960 &  17.4 & 38 \\
CSS0545+02	&  5944.86  & MN & 1200 &  18.9 & 72\tablenotemark{c} \\
CSS0558+00	&  5946.79  & MN &  360 &  16.6 & 7:\tablenotemark{b} \\
CSS0647+49	&  5638     & MT & 23640 &  17.2 & 20 \\
CSS0814$-$00	&  5591     & MT & 18360 &  18.7 & 81 \\
CSS0902$-$11	&  5591     & MT & 23160 &  18.4 & 13 \\
CSS0905+12	&  5589.78  & MT &  1440 & 20.1 & 130 \\
CSS0912$-$03	&  5641     & MT & 19577 &  17.9 & 61 \\
CSS1055+09	&  5587+88  & MT &  2280 &  19.4 & 90 \\
CSS1139+45	&  5589.04  & MT &  1200 & 19.9 & 70: \\
CSS1211$-$08	&  5733.66  & MT &   600 & 17.8 & 79 \\
CSS1556$-$08	&  5592.99  & MT &  1200 & 19.3 & 44 \\
CSS1616$-$18	&  5588.06  & MT &   960 & 15.9 & 2:\tablenotemark{b} \\
CSS1649+03	&  5733.89  & MT &  600 & 18.9 & 97 \\
CSS1706+14	&  5733.8   & MT & 15480 & 17.7 & 84 \\
CSS1720+18	&  5733.91  & MT &  720 & 20.0 & 120 \\
CSS1727+13	&  5734.68  & MT & 1200 & 19.0 & 48 \\
CSS1729+22	&  5739     & MT & 31260 & 18.7 & 60 \\
CSS1735+15	&  5819.65  & OS & 1200 & 18.8 & 17 \\
CSS1752+29	&  5819.71  & OS & 1200 & 17.4 & 16\tablenotemark{b} \\
CSS2029$-$15	&  5819.76  & OS & 1800 & 18.4 & 43 \\   
CSS2059$-$09	&  5820.74  & OS & 1200 & \nodata & 35 \\
CSS2213+17	&  5819.92  & OS & 2160 & \nodata & 80 \\
CSS2227+28	&  5819.89  & OS & 1200 & \nodata & 115 \\
\enddata
\tablecomments{
{\it Column 2:} 
Times are listed as Julian date minus 2 450 000.  Times
for single visits are given to the hundredth; times given
to the nearest day are averages for multi-night observations. 
CSS1055+09 was observed on two nights.  {\it Column 3:} 
OS stands for OSMOS, and M stands for Modspec, with detector
Templeton (T) or Nellie (N).  {\it Column 4:} 
$V$ magnitudes synthesized from the spectrum; they are ideally good to 
$\pm 0.2$ mag, but larger errors are possible because of 
clouds and seeing.  Magnitudes could not be synthesized
for the  spectra of CSS2213+17 and CSS2227+28 because of
condensation on the detector window.  
{\it Column 5:} Positive equivalent widths refer
to emission. 
}
\tablenotetext{a}{CSS0350+35 appears to be a dMe star 
(see text).}
\tablenotetext{b}{These objects appeared to be in outburst,
or are possibly novalike variables.}
\label{tab:spectro_summary}
\end{deluxetable}

\clearpage

\begin{deluxetable}{lrrrrr}
\tablecolumns{6}
\tablewidth{0pt}
\tablecaption{Radial Velocities}
\tablehead{
\colhead{Star} &
\colhead{Time\tablenotemark{a}} &
\colhead{$v_{\rm abs}$} &
\colhead{$\sigma$} & 
\colhead{$v_{\rm emn}$} &
\colhead{$\sigma$} \\
\colhead{} &
\colhead{} &
\colhead{[km s$^{-1}$]} &
\colhead{[km s$^{-1}$]} &
\colhead{[km s$^{-1}$]} &
\colhead{[km s$^{-1}$]} \\
}
\startdata
CSS0501+20 & 55589.6842  &  \nodata & \nodata &  $   86$ & $  18$ \\
CSS0501+20 & 55589.6914  &  \nodata & \nodata &  $   34$ & $  16$ \\
CSS0501+20 & 55589.7017  &  \nodata & \nodata &  $    4$ & $  18$ \\
CSS0501+20 & 55589.7088  &  \nodata & \nodata &  $    5$ & $  19$ \\
CSS0501+20 & 55589.7160  &  \nodata & \nodata &  $   19$ & $  17$ \\
CSS0501+20 & 55589.8673  &  \nodata & \nodata &  $  -10$ & $  15$ \\
CSS0501+20 & 55589.8745  &  \nodata & \nodata &  $   63$ & $  15$ \\
CSS0501+20 & 55589.8817  &  \nodata & \nodata &  $   58$ & $  17$ \\
\enddata
\tablenotetext{a}{Heliocentric Julian Date of mid-exposure minus 2400000; the 
time base is UTC.}
\tablecomments{Table \ref{tab:velocities} is published in its entirety in the electronic 
edition of The Astronomical Journal, A portion is shown here for guidance regarding its form and content.}
\label{tab:velocities}
\end{deluxetable}

\clearpage

\begin{deluxetable}{lllrrcc}
\tablecolumns{7}
\footnotesize
\tablewidth{0pt}
\tablecaption{Fits to Radial Velocities}
\tablehead{
\colhead{Data set} & 
\colhead{$T_0$\tablenotemark{a}} & 
\colhead{$P$} &
\colhead{$K$} & 
\colhead{$\gamma$} & 
\colhead{$N$} &
\colhead{$\sigma$\tablenotemark{b}}  \\ 
\colhead{} & 
\colhead{} &
\colhead{(d)} & 
\colhead{(km s$^{-1}$)} &
\colhead{(km s$^{-1}$)} & 
\colhead{} &
\colhead{(km s$^{-1}$)} \\
}
\startdata

CSS0501+20 emn& 55590.8497(14) & 0.07481(17) &  52(6) & $ 55(4)$ & 48 &  20 \\[1.2ex]
CSS0519+15 emn& 55946.935(2) & 0.0849(3) &  47(7) & $-8(5)$ & 46 &  23 \\[1.2ex]
CSS0647+49 absn & 55641.756(2) & 0.37149(2) &  117(4) & $-51(3)$ & 35 &  13 \\
CSS0647+49 emn & 55641.574(2) & 0.37155(4) &  85(4) & $-87(3)$ & 31 &  11 \\ 
CSS0647+49 comb  & & 0.37150(2) & & & & \\[1.2ex]
CSS0814$-$00 emn & 55591.6527(12) & 0.07485(5) &  61(7) & $ 19(5)$ & 31 &  22 \\[1.2ex]
CSS0902$-$11 absn& 55590.747(3) & 0.2757(4) & 100(6) & $ 42(5)$ & 24 &  19 \\[1.2ex]
CSS0912$-$03 emn & 55640.825(2) & 0.0653(3) &  65(15) & $ 34(10)$ & 29 &  38 \\
 (alternate) & 55640.819(2) & 0.0610(3) &  61(15) & $ 35(10)$ & 29 &  40 \\[1.2ex]
CSS1706+14 emn & 55733.7552(15) & 0.05823(15) &  40(6) & $-46(5)$ & 50 &  24 \\
CSS1729+22 emn & 55738.704(7) & 0.2966(14) &  69(10) & $-11(7)$ & 42 &  35 \\ 
(alternate) & 55738.758(8) & 0.2248(10) &  60(11) & $-22(8)$ & 42 &  40 \\
\enddata
\tablecomments{Parameters of least-squares sinusoid fits to the radial
velocities, of the form $v(t) = \gamma + K \sin[2 \pi(t - T_0)/P]$.}
\tablenotetext{a}{Heliocentric Julian Date minus 2400000.  The epoch is chosen
to be near the center of the time interval covered by the data, and
within one cycle of an actual observation.}
\tablenotetext{b}{RMS residual of the fit.}
\label{tab:parameters}
\end{deluxetable}

\begin{deluxetable}{llrrrr}
\tablecolumns{6} 
\footnotesize
\tablewidth{0pt}
\tablecaption{Standardized Magnitudes}
\tablehead{
\colhead{Object} &
\colhead{Time} &
\colhead{$B-V$} &
\colhead{$V$} &
\colhead{$V-R$} &
\colhead{$V-I$} \\
}
\startdata
CSS0051+20   & 5823.842 &    0.55(5) &  18.66(1) &  \nodata &    1.49(2) \\
CSS0058+28   & 5823.806 &    0.13(9) &  19.38(3) &  \nodata &    0.20(5) \\
CSS0104-03   & 5823.819 &    0.17(18) &  19.80(5) &  \nodata &    0.70(7) \\
CSS0105+11   & 5823.827 &    0.44(23) &  20.36(7) &  \nodata &    0.65(9) \\
CSS0105+19   & 5823.834 &   -0.09(11) &  19.71(4) &  \nodata &    0.07(7) \\
CSS0113+21   & 5823.850 &   -0.12(22) &  20.60(8) &  \nodata &    1.01(10) \\
CSS0116+09   & 5823.856 &    0.24(6) &  18.89(2) &  \nodata &    0.64(3) \\
CSS0150-12   & 5823.864 &    0.02(9) &  19.13(3) &  \nodata &    0.65(4) \\
CSS0208+37   & 5585.586 &    0.22(2) &  18.23(1) &   0.24(1) &   0.38(1) \\
CSS0211+17   & 4524.599 &    0.18(5) &  19.23(3) &  \nodata &    0.90(4) \\
CSS0328+28   & 5585.597 &    0.81(3) &  18.55(1) &   0.40(1) &   0.98(1) \\
CSS0350+35   & 5818.924 &   \nodata &   18.32(2) &  \nodata &   \nodata \\ 
             & 5818.944 &   \nodata &   18.28(2) &  \nodata &   \nodata \\ 
CSS0505+08   & 5820.972 &   \nodata &   18.86(3) &  \nodata &   \nodata \\ 
CSS0545+02   & 5937.729 &    0.47(9) &  18.75(2) &   0.83(2) &   1.71(2) \\
CSS0558+00   & 5937.742 &   \nodata &   20.45(10) &   0.94(11) &   2.09(11) \\
CSS0647+49   & 5585.720 &    0.96(2) &  16.84(0) &   0.66(1) &   1.22(1) \\
CSS0650+41   & 5940.758 &    0.72(13) &  20.07(3) &   0.65(4) &   1.38(5) \\
CSS0814-00   & 5585.812 &   \nodata &   19.35(4) &   0.95(4) &   1.27(4) \\
    & 5585.808 &   \nodata &   19.14(4) &  \nodata &    1.22(5) \\
CSS0821+45   & 5940.842 &   -0.03(17) &  20.52(6) &   0.48(10) &   1.54(7) \\
CSS0845+03   & 4524.703 &   \nodata &   19.91(1) &  \nodata &    0.47(3) \\
CSS1219-19   & 5940.009 &   \nodata &   18.02(3) &   0.31(5) &   0.36(6) \\
   & 5940.020 &    0.49(7) &  18.02(2) &   0.22(4) &   0.20(5) \\
CSS1340+15   & 5940.045 &    0.99(12) &  18.32(1) &   0.69(2) &   1.15(2) \\
CSS1631+10   & 5362.767 &   -0.16(3) &  18.51(2) &  \nodata &    0.81(4) \\
   & 5371.767 &    0.38(11) &  18.19(3) &   0.33(4) &   0.71(4) \\
CSS1639+12   & 5730.810 &   \nodata &   20.52(16) &   0.09(21) &   0.74(20) \\
CSS1644+05   & 5731.796 &   \nodata &   20.36(7) &   0.38(8) &   0.74(8) \\
CSS1649+03   & 5730.816 &    0.50(15) &  18.98(5) &   0.15(7) &   0.51(7) \\
   & 5731.807 &    0.17(7) &  19.01(2) &   0.17(3) &   0.49(3) \\
CSS1701+13   & 5730.841 &   \nodata &   20.85(19) &   0.37(23) &   0.66(26) \\
CSS1720+18   & 5730.833 &   \nodata &   20.46(12) &   0.42(15) &   0.80(16) \\
CSS1727+13   & 5730.824 &    0.29(2) &  17.14(1) &   0.17(1) &   0.38(1) \\
  & 5731.815 &    0.19(2) &  16.96(1) &   0.15(1) &   0.30(1) \\
CSS1752+29   & 5819.705 &   \nodata &   17.34(1) &  \nodata &   \nodata \\ 
\tablebreak
CSS2029-15   & 5732.917 &    0.12(10) &  18.73(4) &   0.18(5) &   0.56(5) \\
   & 5819.750 &   \nodata &   18.05(2) &  \nodata &   \nodata \\ 
   & 5819.751 &   \nodata &   18.19(2) &  \nodata &   \nodata \\ 
CSS2109+13   & 5732.863 &    0.21(2) &  16.56(1) &   0.30(1) &   0.62(1) \\
CSS2109+16   & 5732.850 &   -0.22(12) &  19.45(6) &   0.53(7) &   0.85(7) \\
CSS2144+22   & 5732.872 &    0.50(2) &  17.28(1) &   0.38(1) &   0.72(1) \\
CSS2156+19   & 5732.883 &   -0.96(28) &  20.83(16) &  -0.29(23) &   0.94(18) \\
CSS2158+09   & 5732.891 &   -0.04(3) &  17.48(1) &   0.13(1) &   0.43(2) \\
CSS2213+17   & 5819.899 &   \nodata &   19.69(6) &  \nodata &   \nodata \\ 
CSS2227+28   & 5732.900 &   -0.07(4) &  18.17(1) &   0.53(2) &   1.12(2) \\
             & 5819.876 &   \nodata &   18.76(3) &  \nodata &   \nodata \\ 
CSS2232+18   & 5732.908 &   -0.07(18) &  20.07(7) &   0.47(8) &   0.71(9) \\
CSS2327+41   & 5732.974 &   \nodata &   20.41(22) &   0.42(25) &   0.98(25) \\
\enddata
\tablecomments{Standardized magnitudes and colors of CSS objects.  The 
time given is the Julian date minus 2 450 000.  Statistical uncertainties, in 
hundredths of a magnitude, are given in parentheses.  Where these exceed
$\sim 0.1$ mag, the value should be considered somewhat unreliable.
}
\label{tab:photometry}
\end{deluxetable}

\begin{deluxetable}{lcll}
\tablecolumns{4} 
\footnotesize
\tablewidth{0pt}
\tablecaption{SDSS Classifications and Cross Match}
\tablehead{
\colhead{SDSS CV} &
\colhead{Paper} &
\colhead{CRTS?} &
\colhead{Type} \\
}
\startdata
001856.93+345444.3 &  4 &  No & NL  \\
002603.80-093021.0 &  5 &  No & NCV \\
002728.01-010828.5 &  4 &  No & DN  \\
003941.06+005427.5 &  4 &  No & DN-W \\
004335.14-003729.8 &  3 &  No & DN-W \\
005050.88+000912.7 &  4 &  No & DN  \\
013132.39-090122.3 &  2 &  No & DN-W \\
013701.06-091234.9 &  2 & Yes & DN-W \\
\enddata 
\tablecomments{The SDSS CV sample.  The second
column gives the paper in the SzkodyI-VIII 
series in which the object was published; the third
says whether the object was also found by CRTS; and the 
last gives our estimate of the type as described in the text.
Table \ref{tab:sdssclasses} is published in its entirety in the electronic 
edition of The Astronomical Journal. A portion is shown here for 
guidance regarding its form and content.}
\label{tab:sdssclasses}
\end{deluxetable}

\begin{deluxetable}{lrrr}
\tablecolumns{4} 
\footnotesize
\tablewidth{0pt}
\tablecaption{Catalog Statistics}
\tablehead{
\colhead{Sample} &
\colhead{Total} &
\colhead{Dwarf Novae} &
\colhead{Other Types} \\
}
\startdata
SDSS and CRTS & 44 & 40 & 4 \\
RKcat and CRTS & 119 & 108 & 11 \\
GCVS and CRTS & 34 & 29 & 5 \\
Union & 134 & 123 & 11 \\[2.0ex]
SDSS, not CRTS & 227 & 138 & 89 \\
RKcat, not CRTS & 460 & 250 & 210 \\
GCVS, not CRTS & 253 & 142 & 111 \\
Union & 602 & 348 & 254 \\
\enddata
\tablecomments{Results from cross-matching the CRTS with
other CV samples.  The first three lines give statistics
of objects matched between CRTS and other catalogs, and the
next gives statistics of the union of all these matches;
in this, an object
is counted as a dwarf nova if it is classed as such in 
any of the catalogs.  The bottom four lines repeat this
analysis, but for objects in the catalog that are {\it not}
recovered by CRTS.
}
\label{tab:samplestats}
\end{deluxetable}

\clearpage

\clearpage

\clearpage

\clearpage

\clearpage

\clearpage


\clearpage

\begin{thebibliography}

\bibitem[Aihara et al.(2011)]{sdssviii} Aihara, H., Allende 
Prieto, C., An, D., et al.\ 2011, \apjs, 193, 29 

\bibitem[Archibald et al.(2009)]{archibald} Archibald, A.~M., 
Stairs, I.~H., Ransom, S.~M., et al.\ 2009, Science, 324, 1411 



\bibitem[Baraffe \& Kolb(2000)]{bk00} Baraffe, I., Kolb, U. 2000,
\mnras, 318, 354

\bibitem[Barker \& Kolb(2003)]{barker03} Barker, J., 
\& Kolb, U.\ 2003, \mnras, 340, 623  

\bibitem[Bessell(1990)]{bessell} Bessell, M. S. 1990, \pasp, 102, 1181

\bibitem[Beuermann et al.(1999)]{beuermann99} Beuermann, K., 
Baraffe, I., \& Hauschildt, P.\ 1999, \aap, 348, 524 

\bibitem[Beuermann et al.(1998)]{beuermann2ndry}
Beuermann, K., Baraffe, I., Kolb, U., \& Weichhold, M.\ 1998, \aap, 339,
518


\bibitem[Cash(1979)]{cash79} Cash, W.\ 1979, \apj, 228, 939 



\bibitem[Drake et al.(2009)]{drakecrtts} Drake, A.~J., et al.\ 
2009, \apj, 696, 870 


\bibitem[G{\"a}nsicke et al.(2009)]{unveils} G{\"a}nsicke, 
B.~T., et al.\ 2009, \mnras, 397, 2170 

\bibitem[Horne(1986)]{horne} Horne, K.\ 1986, \pasp, 98, 609 


\bibitem[Kato et al.(2009)]{kato09} Kato, T., et al.\ 2009, 
\pasj, 61, 395 



\bibitem[Keenan \& McNeil(1989)]{keenan89} 
Keenan, P.~C., \& McNeil, R.~C.\ 1989, \apjs, 71, 245 

\bibitem[Samus et al.(2012)]{gcvs} Samus N.N., Durlevich O.V., 
Kazarovets E V., Kireeva N.N., Pastukhova E.N., Zharova A.V., 
et al. General Catalog of Variable Stars 
(GCVS database, Version 2012Jan), Moscow : Sternberg 
Astronomical Institute.  http://www.sai.msu.su/gcvs/gcvs/index.htm 


\bibitem[Knigge(2006)]{kniggedonor} Knigge, C.\ 2006, \mnras, 373,
484

\bibitem[Kolb et al.(1998)]{kolbstillthere} Kolb, U., King, A.~R., 
\& Ritter, H.\ 1998, \mnras, 298, L29 

\bibitem[Kurtz \& Mink(1998)]{kurtzmink} Kurtz, M.~J. \& Mink,
D.~J.\ 1998, \pasp, 110, 934

\bibitem[Landolt(1992)]{landolt92} Landolt, A. U. 1992, AJ, 104, 340


\bibitem[Martini et al.(2011)]{martini11} Martini, P., Stoll, R., 
Derwent, M.~A., et al.\ 2011, \pasp, 123, 187   



\bibitem[Patterson(1998)]{pattlate} Patterson, J.\ 1998, \pasp, 
110, 1132 


\bibitem[Ritter \& Kolb(2003)]{ritterkolb} 
Ritter, H., \& Kolb, U.\ 2003, \aap, 404, 301 

\bibitem[Schlegel, Finkbeiner, \& Davis(1998)]{schlegel98}
Schlegel, D. J., Finkbeiner, D. P., \& Davis, M. 1998, \apj, 500, 525

\bibitem[Schneider \& Young(1980)]{sy80} Schneider, D. and Young, P. 1980,
\apj, 238, 946

\bibitem[Shafter(1983)]{shafter83} Shafter, A.~W.\ 1983, \apj,
267, 222




\bibitem[Szkody et al.(2002)]{szkodyi} Szkody, P., et al.\ 
2002, \aj, 123, 430 

\bibitem[Szkody et al.(2003)]{szkodyii} Szkody, P., et al.\ 
2003, \aj, 126, 1499 

\bibitem[Szkody et al.(2004)]{szkodyiii} Szkody, P., et al.\ 
2004, \aj, 128, 1882 

\bibitem[Szkody et al.(2005)]{szkodyiv} Szkody, P., et al.\ 
2005, \aj, 129, 2386 

\bibitem[Szkody et al.(2006)]{szkodyv} Szkody, P., et al.\ 
2006, \aj, 131, 973 

\bibitem[Szkody et al.(2007)]{szkodyvi} Szkody, P., et al.\ 
2007, \aj, 134, 185 

\bibitem[Szkody et al.(2009)]{szkodyvii} Szkody, P., et al.\ 
2009, \aj, 137, 4011 

\bibitem[Szkody et al.(2011)]{szkodyviii} Szkody, P., Anderson, 
S.~F., Brooks, K., et al.\ 2011, \aj, 142, 181 

\bibitem[Thorstensen \& Freed(1985)]{tf85} Thorstensen, J.~R., \& Freed, I.~W.\ 1985, \aj, 90, 2082 


\bibitem[Thorstensen et al.(1996)]{tpst} Thorstensen, J. R., Patterson, J.,
Thomas, G., \& Shambrook, A. 1996, \pasp, 108, 73

\bibitem[Thorstensen et al.(2002)]{thorstenseneiuma} Thorstensen, J.~R., 
Fenton, W.~H., Patterson, J.~O., et al.\ 2002, \apjl, 567, L49 




\bibitem[Thorstensen \& Armstrong(2005)]{thorj1023} Thorstensen, J.~R., 
\& Armstrong, E.\ 2005, \aj, 130, 759 


\bibitem[Tonry \& Davis(1979)]{tondav} Tonry, J.~\& Davis, M.\
1979, \aj, 84, 1511

\bibitem[Unda-Sanzana et al.(2008)]{undana-gd552} Unda-Sanzana, E., 
Marsh, T.~R., G{\"a}nsicke, B.~T., et al.\ 2008, \mnras, 388, 889 


\bibitem[Wang et al.(2009)]{wang} Wang, Z., Archibald, 
A.~M., Thorstensen, J.~R., et al.\ 2009, \apj, 703, 2017 

\bibitem[Warner(1995)]{warner95} Warner, B., in {\it Cataclysmic Variable
Stars}, 1995, Cambridge University Press, New York

\bibitem[Warner \& Woudt(2010)]{ww10} Warner, B., \& Woudt, P.~A.\ 2010, 
American Institute of Physics Conference Series, 1314, 185 

\bibitem[Witham et al.(2007)]{withamiphas} Witham, A.~R., Knigge, 
C., Aungwerojwit, A., et al.\ 2007, \mnras, 382, 1158 


\bibitem[Woudt et al.(2012)]{ww12} Woudt, P.~A., Warner, B., 
de Bud{\'e}, D., et al.\ 2012, \mnras, 2533 


\end{thebibliography}
\end{document}